\documentclass{article}
\usepackage{arxiv}
\usepackage{hyperref}
\usepackage[utf8]{inputenc} 
\usepackage[T1]{fontenc}    
\usepackage{booktabs}       
\usepackage{nicefrac}       
\usepackage{microtype}      
\usepackage{graphicx}
\usepackage{natbib}
\usepackage{doi}
\usepackage{algorithm}
\usepackage{algpseudocode}
\usepackage{multirow}
\usepackage{booktabs}
\usepackage{pdflscape}
\usepackage{rotating}
\usepackage{enumitem}
\usepackage{caption}
\usepackage{xurl}
\usepackage[caption=false,font=normalsize,labelfont=sf,textfont=sf]{subfig}
\usepackage[normalem]{ulem}
\usepackage[table]{xcolor}

\title{Heuristics for AI-driven Graphical Asset Generation Tools in Game Design and Development Pipelines: A User-Centred Approach}

\date{June 27, 2025}

\author{\href{https://orcid.org/0000-0001-9828-7641}{\includegraphics[scale=0.06]{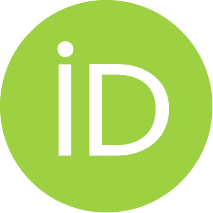}\hspace{1mm} Kaisei Fukaya}, \href{https://orcid.org/0000-0001-7849-458X}{\includegraphics[scale=0.06]{orcid.pdf}\hspace{1mm}Damon Daylamani-Zad}, \href{https://orcid.org/0000-0002-8818-2683}{\includegraphics[scale=0.06]{orcid.pdf}\hspace{1mm}Harry Agius} \\
\texttt{\{kaisei.fukaya, damon.daylamani-zad, harry.agius\}@brunel.ac.uk}\\
College of Engineering, Design and Physical Sciences\\
Brunel University of London\\
Uxbridge, Middelsex, UK\\
}

\renewcommand{\shorttitle}{Heuristics for AI-driven Graphical Asset Generation Tools in Game Dev Pipelines}

\begin{document}
\maketitle

\begin{abstract}
Graphical assets play an important role in the design and development of games. There is potential in the use of AI-driven generative tools, to aid in creating graphical assets, thus improving game design and development pipelines. However, there is little research to address how the generative methods can fit into the wider pipeline. There also no guidelines or heuristics for creating such tools. To address this gap we conducted a user study with 16 game designers and developers to examine their behaviour and interaction with generative tools for graphical assets. The findings highlight that early design stage is preferred by all participants. Designers and developers are inclined to use such tools for creating large amounts of variations at the cost of quality as they can improve the quality of the artefacts once they generate a suitable asset. The results also strongly raised the need for better integration of such tools in existing design and development environments and the need for the outputs to be in common data formats, to be manipulatable and smoothly integrate into existing environments. The study also highlights the requirement for further emphasis on the needs of the users to incorporate these tools effectively in existing pipelines. Informed by these results, we provide a set of heuristics for creating tools that meet the expectations and needs of game designers and developers.
\end{abstract}

\keywords{Graphical game assets, Artificial Intelligence, PCG, User preference, User interface}

\section{Introduction}

While usage of procedural content generation (PCG) in games is widespread, it is disproportionately applied to certain forms of content,  such as environments, levels and narratives. Games such as No Man's Sky \citep{HelloGames2016}, Minecraft \citep{MojangStudios2011} and Dwarf Fortress \citep{Bay12Games2006} use PCG to create new content for  players to see and explore, while others use it to streamline development, such as Starfield \citep{skrebels2023starfield} in which large quantities of content are generated then refined by hand.

Togelius et al. \citep{Togelius2011} introduce the distinction between \textit{online} and \textit{offline} forms of PCG. \textit{Online} PCG occurs during the game's runtime, while \textit{offline} PCG occurs during the game's development. In \textit{online} PCG, care must be taken to ensure that the range of outcomes are within the designers intentions, this naturally limits the forms of content for which PCG can be applied in a practical sense. Most cases will therefore focus on types of content such as level-layouts, or item stats, which can easily be constrained and balanced in the interest of players. \textit{Offline} PCG affords the system more freedom, as all content is curated by a human designer. This leads to further possibilities in providing designers and developers more creative control over the content produced this way. Mixed-Initiative Procedural Content Generation (MI-PCG), involves both humans and PCG systems in an iterative design process \citep{Liapis2016, lai2020towards}.

It is apparent that different generative methods and techniques are designed with different use cases in mind, each attempting to solve a problem or streamline a process within a larger creative pipeline. Practitioners use many tools during game design and production, from editor tools for game engines, to asset editing software such as Photoshop and Blender, or integrated development environments such as Visual Studio. These tools each have their place within the workflows and development pipelines of game creators. While these workflows look different from company to company or from one individual to another, there exists three ubiquitous stages of development: prototyping/design, production and testing \citep{Ramadan2013}. All graphical assets are formulated and produced during the prototyping, design and production stages. Within this, there are many ways that PCG can help, from early inspiration and creating rough placeholders to creating fully fledged assets or remixing existing assets for variety. Each of these use cases, naturally, will have different requirements for the complexity of input, method of interaction and quality of output. However, no research has yet explored the behaviour and interaction patterns of the users, i.e. the purposes for which designers and developers use PCG tools and how they fit in their pipelines.

While there is an extensive body of literature pertaining to MI-PCG there are limited case studies which examine their use and fit in design and development pipelines \citep{lai2020towards}. While it is clear that MI-PCG has the potential to improve game development asset pipelines, the behaviour, opinions and preferences of game designers and developers remain, in general, unheard. Though recent work examines usage applications for generative AI systems such as ChatGPT \citep{YAN2024103867} among general user bases, the specific needs and requirements within the video game industry and research have not yet been addressed. Individual MI-PCG systems may be validated with user feedback \citep{Li2021d, shen2020deepsketchhair, shen2021clipgen}, but feedback is limited to specific purposes of use, and are not contextualised within the larger scope of the development pipeline. The goal of this study is to obtain insights into where and how graphical asset PCG is most useful to game designers and developers.

The following section, explores the current state of research in this area. Section \ref{sec:motivation} will expand on the motivation for this study and identify seven pipeline applications. Section \ref{sec:researchQ} will then introduce the research questions. Section \ref{sec:procedure} will describe the study procedure, followed by sections \ref{sec:mockups} and \ref{sec:results} which will explain the mock-ups that were tested and the experiment results respectively. Sections \ref{sec:comparison} and \ref{sec:heuristics} will discuss the findings and present our heuristics for creators of graphical asset generative tools, and further research.

\section{Review of Current Research}\label{sec:litreview}
The process of creating games is complex, requiring design decisions at every level from the overall concept and gameplay down to the characters and environments. Designers and developers make use of increasingly capable creative tools to streamline this task, which they often tailor to their own needs \citep{9799134, seidel2019designing}.

In a qualitative study examining expectations of Finnish game developers regarding their tools, Kasurinen et al. \citep{kasurinen2013game} present some key insights. Across the seven organisations examined, developers were largely satisfied with the tools they have, though, when it comes to assets, many preferred to purchase rather than create in-house. It is found that the companies 'expect their tools to allow easy prototyping and the ability to design while implementing' \citep{kasurinen2013game}. Many of the companies relied on third-party game engines, and thus compatibility with these engines was a large part of the consideration when selecting new tools. 

The vast amount of content required in video games requires an equally vast amount of time and resources to create. Generative approaches can be applied to reduce this work load \citep{seidel2019designing}.

It is important for designers and developers to have control over the content that is created, thus generative tools typically involve mixed-initiative (MI) approaches. The user study of Walton et al. \citep{walton2021evaluating}, examines user opinions of a PCG level creation tool using a mixed-methods approach. The findings suggest that the MI-PCG method stimulates user creativity, and helps to inspire new ideas. They discover that the inclusion of qualitative data from participants is important for gaining the full picture, as it provides important context to the quantitative results.

The concept of designer modeling extends MI-PCG by considering the adaptability of a system to user needs \citep{liapis2013designer}. In such a system, the tool aligns itself to the designers style choices, or may even help break design fixation by producing results that are different to what the user is accustomed to, providing inspiration.

Sketchar \citep{Ling2024sketchar} is a mixed-initiative generative tool encompassing a full character design pipeline. In this method, users iteratively build on each aspect of a design, including character profiles, concept art, relationships with other characters and dialogue between them, using large language models (LLMs) such as ChatGPT and image generators such as DALL-E. In user testing, designers identify the benefits of this method in elaborating on existing, early-stage ideas as well as visually communicating their design ideas to artists.

Colado et al. \citep{colado2023using}, examine mixed-initiative generative approaches to serious game design. Much like Sketchar \citep{Ling2024sketchar}, this method implemented an iterative text-to-image generation workflow, this time for character concepts as well as scenarios. Results from initial testing suggests that this method can save time in development and aid smaller teams with less resources.

It is clear that generative methods have clear applications in graphical asset pipelines. According to Unity's game report in 2024, 62\% of developers that have adopted AI tools use them for asset generation \citep{unity2024}. However, the impact of generative, mixed-initiative tools for graphical game assets is yet to be examined.

\section{Motivation \& Current Development Pipeline}\label{sec:motivation}

\begin{figure*}[!t] 
\centering
\includegraphics[width=\linewidth]{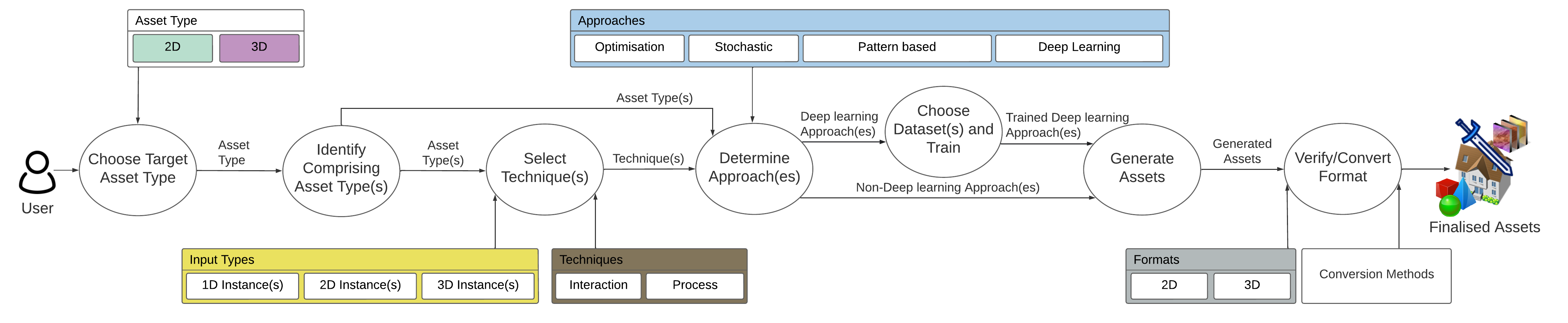}
\caption{The \textit{Graphical Asset Generation/Transformation} (GAGeTx) framework. A framework for designing graphical asset generators based on user needs and requirements.}\label{fig:GaGeTx}
\end{figure*}

\textit{Graphical Asset Generation/Transformation} (GAGeTx) is a framework for designing graphical asset generators (GAGs) based on user needs and requirements \citep{fukaya2023intelligent, fukaya2024evaluation}. It tasks the user with making a series of decisions that feed into one-another, including the type of asset to be generated, the technique and approach to be used, and whether data format conversion is required as part of the pipeline. The high level summary of GAGeTx is presented in Figure \ref{fig:GaGeTx}.

This framework currently addresses the \textit{how} of graphical asset generation. The user chooses an asset type and dimensionality, selects a technique to use and considers the input types required for the technique. A relevant approach can then be selected based on whether it uses the chosen technique and input types, and generates the target asset type. Format conversion may then be applied if the chosen approach does not produce results in the format required by the user. This, however, begins with the assumption that the user knows what they need to generate, how they will incorporate it within their pipeline, and the quality to be expected.

However, the framework does not yet consider the issues of \textit{where to generate} and \textit{how to interact}. In a practical scenario, a generative tool for creating assets will likely function within a larger pipeline. There should be a clear purpose for the generative tool. Consideration for the purpose of a generator also has implications on the standard to which it is implemented, the compromise between quality of output, and the volume or variety of results. For example, a generator used for early prototyping or ideation in the pipeline must meet different standards of quality and speed to one used for producing 'game ready' assets, which is a stage toward the end of the pipeline. This position in the development pipeline may also dictate the input types available, their quality, and the output format needed. Which also has an affect on the choice of approach. Furthermore, how the tool is interacted with and the provided features should be considered, as well as how this is presented to the user. This ties in with the purpose, as for example, fine-grained control can be a requirement for 'game ready' asset creation, but a hindrance in the ideation phase where maximal speed and minimal interaction is required.

Through the analysis of \textit{\textbf{existing game development pipelines}}, Ramadan and Widyani, \citep{Ramadan2013} introduced a comprehensive game development life cycle (GDLC). In this framework, practitioners begin with a rough concept of the game then begin a design and development loop starting with \textit{pre-production}, in which more detailed concepts are established, then followed by \textit{production}, in which the concepts are implemented, and \textit{testing} in which these implementations are appraised. Insights from testing then feed into further pre-production and the cycle continues, iterating on the design until the practitioners are satisfied and they move on to refinement in the beta phase, then finally release the product.

\begin{figure*}[!t] 
\centering
\includegraphics[width=1\linewidth]{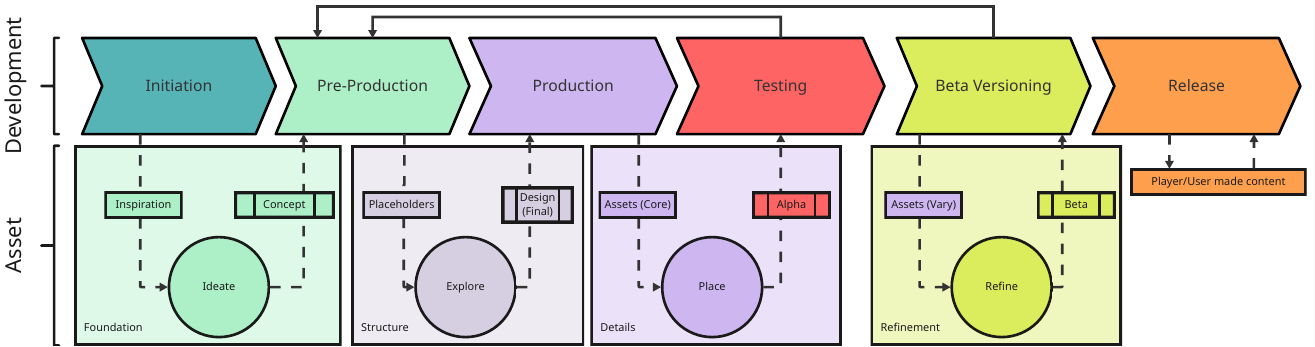}
\caption{Current game design and development pipeline, adapted from \citep{Ramadan2013}.}\label{fig:initdevpipeline}
\end{figure*}

To build on this, we present an expanded version of the GDLC, incorporating the graphical asset creation pipeline as it relates to general development, shown in figure \ref{fig:initdevpipeline}. Here we identify points in the pipeline at which practitioners utilise creative tools, through observing existing generative methods, tools and pipelines in the literature. These include building inspiration in the interim between initiation and pre-production; generating placeholder assets, and exploring designs during pre-production and early production; implementing these designs as full (Core) assets in production; creating variations of these assets in the refinement stage; and supporting player/user made content. 

\begin{figure}[!t] 
\centering
\includegraphics[width=.75\linewidth]{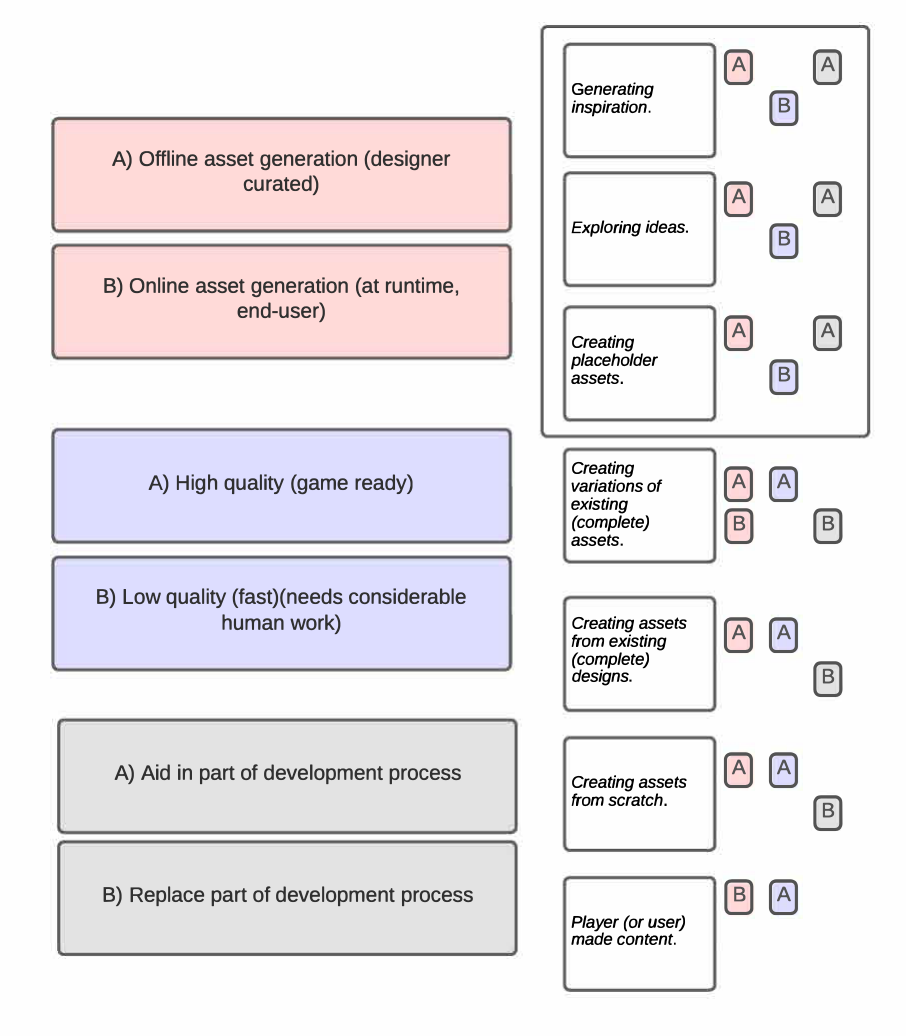}
\caption{The key characteristics of design and production uses. \textit{Generating inspiration}, \textit{Exploring ideas} and \textit{Creating placeholder assets} are grouped together as they share the same characteristics.}\label{fig:pipelinecat}
\end{figure}

\textbf{Generating inspiration:}
Creative block and design fixation \citep{JANSSON19913} are widely experienced in creative fields, including game production. Generative methods for graphical assets can be used to help circumvent this issue by providing unexpected or varied outputs that can serve as inspiration. In this context, the quality of the generated content is less important than its ability to spark the user's creativity, serving as a starting point or trigger for further design iterations.

\textbf{Exploring ideas:}
During early stages of design, iteration is key for expanding on initial inspirations and producing finalised designs. During this process, designs are flexible, thus ideas can be quickly tested and evaluated before committing extensive resources and time. The speed and configurability of some generative methods such as \citep{Saharia2022} may help with exploring ideas and potential design directions for this use case.

\textbf{Creating placeholder assets:}
Placeholder graphics are assets that are used as a stand-in for future intended game elements. They are used during the prototyping phase to allow for the testing of game features and mechanics. As the intention is for these assets to be replaced during later stages, quality and detail is unnecessary. However, to serve fast prototyping, placeholder assets must be fast to produce.

\textbf{Creating variations of existing (complete) assets:}
In games, asset variation can help to immerse players and reduce the repetitiveness of content throughout a digital environment. Many generative methods facilitate the ability to create similar variations of assets. This is most commonly seen with methods that rely on parametric modelling \citep{Jones2021, Getto2020}, or with the usage of VAEs \citep{tan2018variational}. Therefore, there is potential in the usage of generative tools for increasing the diversity and richness of content available, making the digital environment more engaging and varied for players.

\textbf{Creating assets from existing (complete) designs:}
During the design phase, the appearance of an asset may be determined and planned through the use of concept art. This concept art is later used as reference for creating the final asset during the production phase. A graphical asset generator may be used to achieve the latter task, by converting initial designs and plans into finalised assets. For example, \citep{SilvaJunior2018} use orthographic concept art from multiple views to generate 3D assets, even utilising the initial art for texturing.

\textbf{Creating assets from scratch:}
Some generative tools, such as SpeedTree \citep{InteractiveDataVisualizationInc.IDV2022}, encompass a full pipeline for designing, configuring then producing assets. In these cases, the process covers the creation of assets from scratch.

\textbf{Player (or user) made content:}
A common application of PCG, beyond its use in generating graphical assets, is for \textit{online} content generation. This typically refers to the generation of game elements such as levels and loot to produce unexpected and re-playable experiences. With regard to graphical assets however, some games such as Spore and Dreams provide tools that allow players to create their own designs within a constrained creation framework. In a sense, many of these tools are constrained versions of development tools, there is therefore potential for the application of generative tooling within similar systems. Generative systems can also be used for avatar personalisation, such as through the reconstruction of faces from photographs \citep{Lin2021b}.

Figure\ref{fig:pipelinecat} presents the key characteristics of each game centred use case. As shown, graphical asset generators may be applied \textit{offline} during the production of a game product, or \textit{online} during run-time. Uses may also necessitate high quality outputs, such as in \textit{creating variations of existing designs} and \textit{creating assets from scratch}, or low quality outputs with an emphasis on speed, such as in the case of \textit{generating inspiration}. These uses may also aid in streamlining part of a development process, such as with \textit{exploring ideas}, or replace an entire process, as with \textit{creating assets from scratch}.

These characteristics represent the underlying acceptable qualities of the usages, helping to map different generative methods to specific stages in the development pipeline. For instance, during the early design phase, a generator prioritising speed over quality can help by rapidly generating ideas, thereby encouraging creative exploration. Alternatively, high-quality generators are more suited for final asset production, where precision and detail is necessary.

\section{Research Questions}\label{sec:researchQ}

The aim of this study is two-fold: a) to establish if such generative tools would be found useful by game designers and developers, and what their requirements are for interacting with such tools, and b) to discover where in the pipeline such tools would best fit, and how the chosen generative method may line up with this. As a result, this will provide expand and provide nuance to the GAGeTx framework.

To achieve this, this study will obtain game designer and developer feedback with regard to the concept of generating assets for game projects, guided by visual interaction with UI mock-ups. To ensure that opinions are not shaped by the type of asset, technique or approach to generation; the UI mock-ups are designed to present a fully customisable generative system, where the user decides these specifics.

There are five core questions this study seeks to answer, shown below in order of importance.

\begin{enumerate}\label{tab:coreQuestions}
    \item[CQ 1] Would a generative system for asset generation be useful to game designers/developers?
    \item[CQ 2] Where in the design/development pipeline would game designers/developers find value in such as system?
    \item[CQ 3] What are the expectations regarding speed and quality for such a system?
    \item[CQ 4] Is this type of system preferred as integrated or stand-alone?
    \item[CQ 5] Which type/s of UI interaction do game designers/developers prefer for this type of system?
\end{enumerate}

A mixed methods approach is taken to collect quantitative and qualitative data from participants. This methodology attempts to answer a series of research questions, which expand the above; presented in table \ref{tab:researchQuestions} as RQ1-RQ12, by proving or disproving a relative set of hypotheses, listed H1-H7 in table \ref{tab:mainHypotheses} and demographical impact hypotheses RH1-RH3 listed below. To achieve this, three prototypes have been devised, as shown in figure \ref{fig:mockupSteps}.

\begin{table*}[h]
\caption{List of research questions and associated questionnaire questions.}
\label{tab:researchQuestions}
\begin{tabular}{lp{.4\linewidth}p{.4\linewidth}}
\multicolumn{2}{c}{Research Questions} & Survey Questions\\ \toprule
RQ1                & Which prototype/mock-up (\textit{M\textsubscript{i}}) is deemed the most useful to game designers/developers?& "I would find this software tool that generates graphical assets useful in the projects I work on."\\ \midrule
RQ2                & Where in the pipeline would game designers/developers find value in \textit{M\textsubscript{i}}?& "Where in your development pipeline would you find value in this software tool?"\\
\hspace{3mm} RQ2.1              & Do designers/developers find value in \textit{M\textsubscript{i}} for Generating inspiration.\\
\hspace{3mm} RQ2.2              & Do designers/developers find value in \textit{M\textsubscript{i}} for Exploring ideas.\\
\hspace{3mm} RQ2.3              & Do designers/developers find value in \textit{M\textsubscript{i}} for Creating placeholder assets.\\
\hspace{3mm} RQ2.4              & Do designers/developers find value in \textit{M\textsubscript{i}} for Creating variations of existing (complete) assets.\\
\hspace{3mm} RQ2.5              & Do designers/developers find value in \textit{M\textsubscript{i}} for Creating assets from existing (complete) designs.\\
\hspace{3mm} RQ2.6              & Do designers/developers find value in \textit{M\textsubscript{i}} for Creating assets from scratch.\\
\hspace{3mm} RQ2.7              & Do designers/developers find value in \textit{M\textsubscript{i}} for Player (or user) made content.\\ \midrule
RQ3                & If \textit{M\textsubscript{i}} were to be used in a pipeline, would designers/developers prioritise volume of output or quality of output?& "Considering your answer to the previous questions, would you prefer if this tool generated a large variety of assets at a lower quality, or that it generated a small variety of high-quality assets?"\\ \midrule
RQ4                & How much time would designers/developers find acceptable for \textit{M\textsubscript{i}} to take in generating a single asset?& "Please indicate the largest timescale you would find acceptable for generating a single graphical asset, if you were to use this software tool in your projects."\\ \midrule
RQ5                & Do designers/developers prefer \textit{M\textsubscript{i}} as a stand-alone solution or integrated into a game-engine?& "Given the option, would you prefer such a system to exist as stand-alone software, or integrated into your game engine editor of choice?"\\ \midrule
RQ6                & How important is the ability to configure or modify \textit{M\textsubscript{i}} according to designers/developers?& "Please rate on a scale of 1 (not important) to 5 (very important), how important to you, is the ability to configure or modify such a tool. For example, importing your own bespoke algorithms or trained models?"\\ \midrule
RQ7                & Which \textit{M} do designers/developers identify as more useable? & System Usability Scale \citep{brooke1996sus}\\ \midrule

RQ8               & What are the concerns and perceived benefits of this tool? \\ \midrule
RQ9              & What are the desired asset types? \\ \bottomrule                                            
\end{tabular}
\end{table*}

\begin{table}[]\caption{List of hypotheses and associated research questions.}\label{tab:mainHypotheses}
\centering
\begin{tabular}{lp{55mm}l}
\multicolumn{2}{l}{Hypotheses}  & Associated RQs\\ \toprule
H1& Participants prefer an integrated solution over a standalone& RQ1, RQ5\\\midrule
H2& Participants favour the integrated window interface type over an inspector integrated interface& RQ1, RQ6.1\\\midrule
H3& Participants find the tool valuable in all stages& RQ2\\\midrule
H4& Participants prefer shorter generation times over quality/variety&RQ3\\\midrule
H5& Participants find asset generation times of less than a minute acceptable &RQ4\\\midrule
H6& Participants prefer an open solution i.e., high-configurability& RQ6\\\midrule
H7& Participants identify the solutions useable& RQ7\\ \bottomrule
\end{tabular}
\end{table}

\begin{enumerate}\label{tab:impactHypotheses}
    \item[RH1.1] Role impacts preference.
    \item[RH1.2] Size of team impacts preference.
    \item[RH2.1] Experience impacts preference.
    \item[RH2.2] Age impacts preference.
    \item[RH3.1] Gender impacts preference.
\end{enumerate}

\section{Procedure}\label{sec:procedure}
To obtain preferences of game developers and designers with regard to generative tools for graphical assets, we developed three mock-up interfaces, designed to cover broad forms of interaction. Each participant was asked to complete a form consisting of demographic questions, followed by a section for each mock-up. In each section the mock-up is introduced and instructions for installing and using the mock-up are provided. The participant is asked to try the mock-up for as long as they like, then complete a set of preference questions. After all three sections are completed, the participant is provided with an opportunity to book a semi-structured interview to provide qualitative feedback. In analysis, the questionnaire responses are compared between the three mock-ups, and overall. The study was approved by University's Research Ethics Committee (37113-LR-Jun/2022-39654-1). The following subsections will provide more detail on the participants, mock-ups, questionnaires and interviews.

\subsection{Participants}
 Participants were recruited via a convenience sampling approach, in which members of the game development communities: LUUG and BCS Animation and Games specialist group, were contacted via group emails through the respective community administrators. Participants were required to be over the age of 18, with professional experience in game design or development. Participants were also required to have Unity engine installed on their personal machines, and experience with the software was preferred. Of the group email respondents, participants were recruited based on the aforementioned criteria, and signed a consent form before taking part the study. Participation was not incentivised on the basis of financial gain, and was thus voluntary.

The pool of participants was formed of 16 (13 male, 3 female) game designers and developers, aged 18-39 years old. As the exact age of users was not required, they were asked to choose their age range, 81\% were in 18-25 category and 19\% in 35-39 category. Ten participants had one year professional experience in the industry, three had between 1-4 years of professional experience and three had 5-9 years of experience. The minimum team size reported was one and the maximum current team size of the participants was 25, with a median of 11. Exactly half of the participants (50\%) reported as artists and designers, and the other half (50\%) reported working as developers which included technical artists and programmers.

\subsection{Interface prototype mock-ups}
In order to explore the preferences of participants, three prototype mock-ups have been devised, as presented in section \ref{sec:mockups}. These mock-ups M\textsubscript{1}, M\textsubscript{2} and M\textsubscript{3} represent three forms of tool implementation; stand-alone, integrated in an editor inspector and integrated in an editor window respectively. These mock-ups were designed as  representations of the flow of interaction and were therefore not functional under the hood. M\textsubscript{2} and M\textsubscript{3} have been implemented in the Unity engine. Figure \ref{fig:mockupSteps} shows a step-by-step process within each of the three mock-ups, achieving an equivalent result. As the figure shows, \textit{M\textsubscript{1}} takes a direct approach by asking the user a series of questions that drill down on a configuration based on the user's need. \textit{M\textsubscript{2}} and \textit{M\textsubscript{3}} take a more free-form approach, giving the user a palette of options from which to build their generator. Users were provided walk-through videos to watch before using each interface, explaining their usage and the features available.

\subsection{Questionnaires}
Participants were given the opportunity to explore each mock-up at their own discretion and complete a questionnaire for each one, as well as a fourth questionnaire for collecting data independent of the mock-ups, such as demographic data. These questionnaires consist of 5-point Likert scales, multiple-choice tick boxes and dichotomous questions. Demographic data, such as age, years of experience and team size, were recorded by range. These surveys were hosted on Microsoft Forms. The research questions and the corresponding survey questions are presented in table \ref{tab:researchQuestions}.

\subsection{Semi-structured interviews}
Participants were scheduled 10-to-15-minute interviews following their completion of the previous step, this was on an opt-in basis. In these interviews, participants were first asked questions to confirm their answers within the questionnaires to ensure validity, then asked to describe any concerns or perceived benefits of such a tool, what asset types they desire to generate, and what additional features they desire. Participants were also given the opportunity to voice any opinions that had not been addressed at any other stage in the study. Notes were taken during these semi-structured interviews for later analysis.

\section{Mock-ups}\label{sec:mockups}
In order to obtain game designer and developer feedback, three UI mock-ups have been developed and compared. To ensure that opinions are not shaped by the type of asset, technique or approach to generation; the UI mock-ups are designed to present a fully customisable generative system, where the user decides these specifics.

Each graphical asset generator can consist of any number of techniques, which are in turn comprised of an \textit{interaction} type and a \textit{process} type. Inputs are provided via the interaction type and outputs of the process can be fed as an input to another technique, or as a final artefact. In addition, a user can choose to convert the format of input and output data as needed.

\begin{figure*}[ht] \caption{Equivalent step-by-step process for using each interface.} \label{fig:mockupSteps}
\centering
\minipage{0.195\textwidth}
    \centering
  \includegraphics[width=\linewidth]{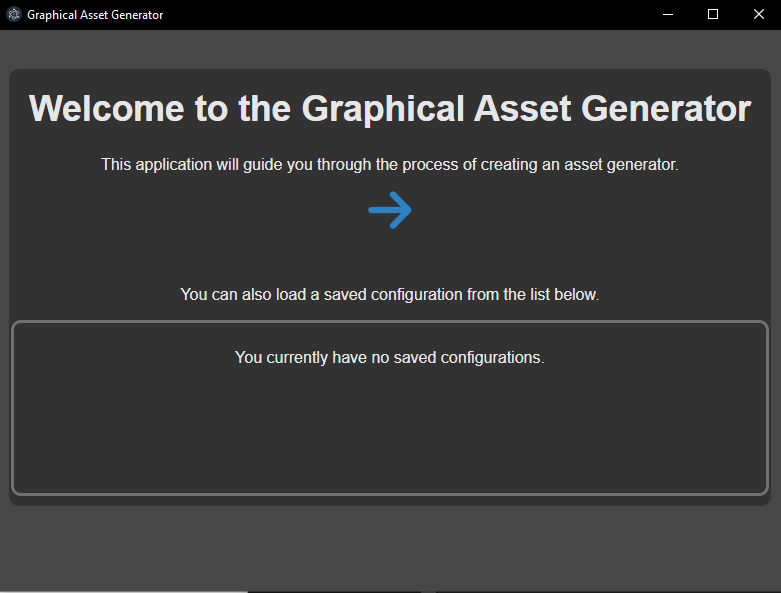}
  \\ Creating a new graphical asset generator 
\endminipage\hfill
\minipage{0.195\textwidth}
    \centering
  \includegraphics[width=\linewidth]{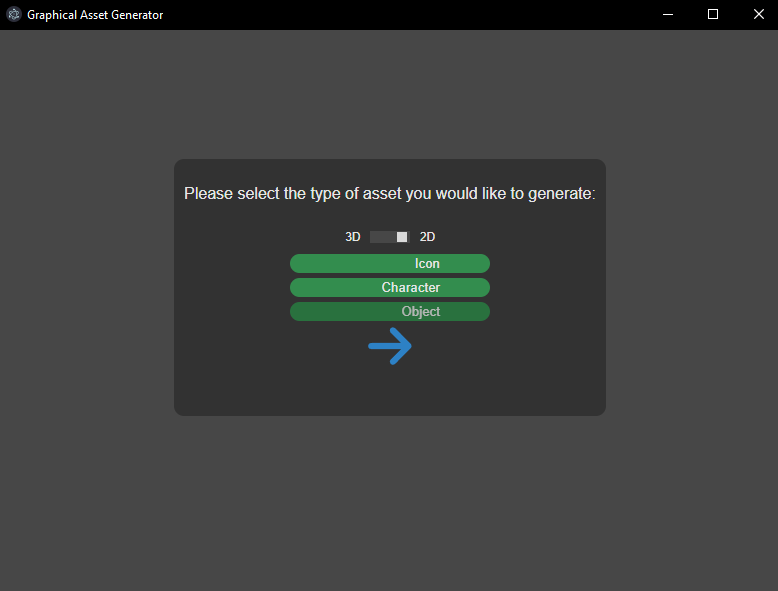}
  \\ Defining target asset type 
  \vspace{.4cm}
\endminipage\hfill
\minipage{0.195\textwidth}
\centering
  \includegraphics[width=\linewidth]{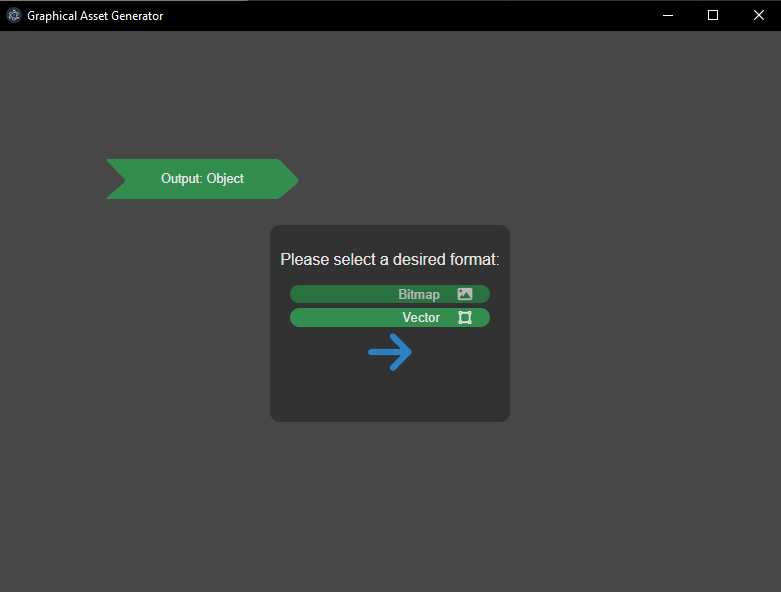}
    \\ Defining target format 
    \vspace{.8cm}
\endminipage\hfill
\minipage{0.195\textwidth}
\centering
  \includegraphics[width=\linewidth]{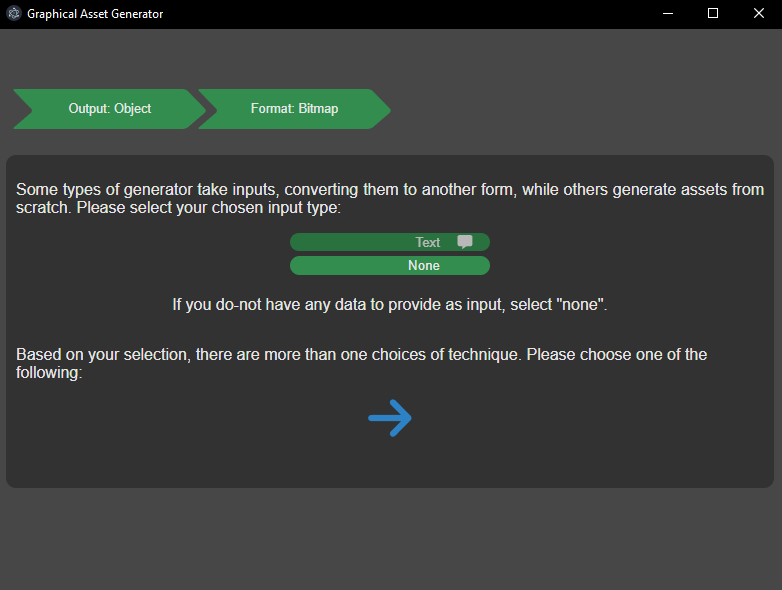}
    \\ Assigning the technique 
    \vspace{.4cm}
\endminipage\hfill
\minipage{0.195\textwidth}
\centering
  \includegraphics[width=\linewidth]{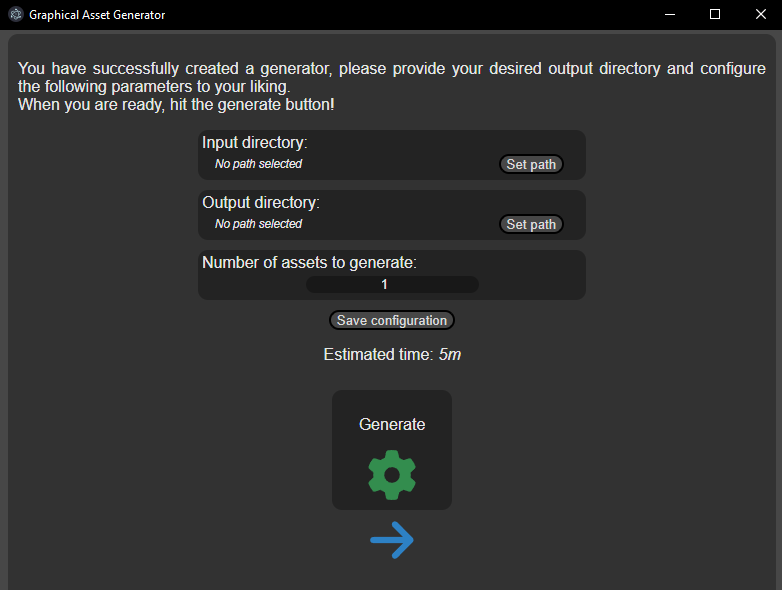}
    \\ Assigning input and output sources 
    \vspace{.4cm}
\endminipage
\vspace{.25cm}
\minipage{\textwidth}
\centering
\textit{Stand-alone wizard interface (M\textsubscript{1})}
\endminipage
\vspace{.3cm}

\minipage{0.245\textwidth}
\centering
  \includegraphics[width=\linewidth]{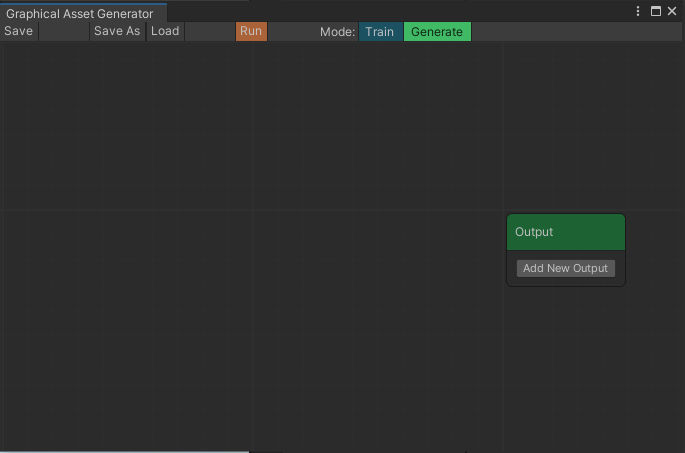}
  \\ Creating a new empty graphical asset generator graph 
\endminipage\hfill
\minipage{0.245\textwidth}
\centering
  \includegraphics[width=\linewidth]{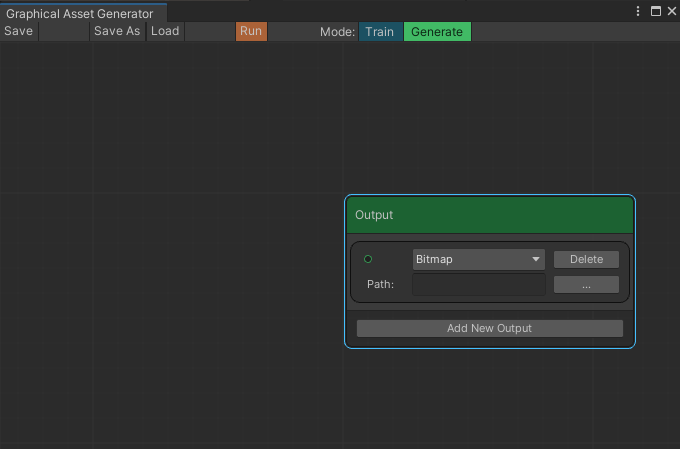}
    \\ Defining target asset type and format 
    \vspace{.4cm}
\endminipage\hfill
\minipage{0.245\textwidth}
\centering
  \includegraphics[width=\linewidth]{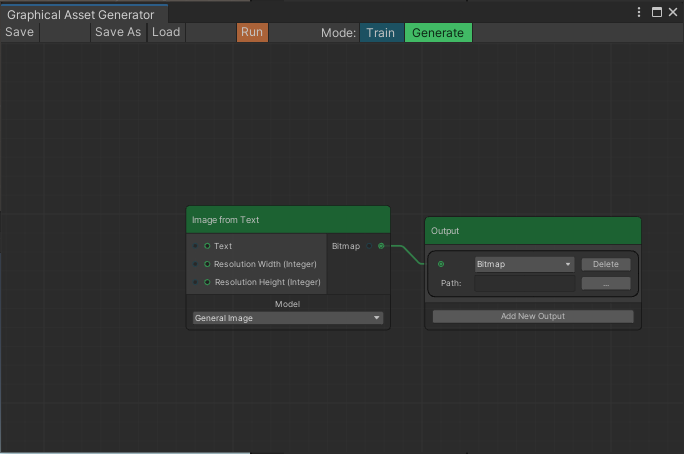}
    \\ Assigning the technique 
    \vspace{.8cm}
\endminipage\hfill
\minipage{0.245\textwidth}
\centering
  \includegraphics[width=\linewidth]{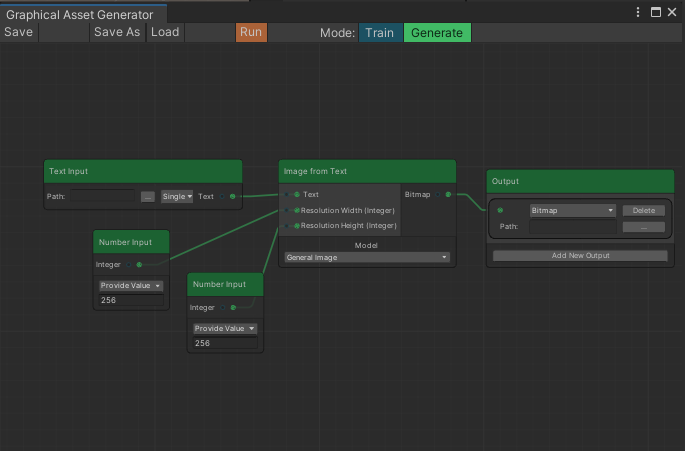}
    \\ Assigning input and output sources 
    \vspace{.4cm}
\endminipage
\vspace{.25cm}
\minipage{\textwidth}
\centering
\textit{Integrated in an editor window interface (M\textsubscript{2})}
\endminipage
\vspace{.3cm}

\minipage{0.195\textwidth}
\centering
  \includegraphics[width=\linewidth]{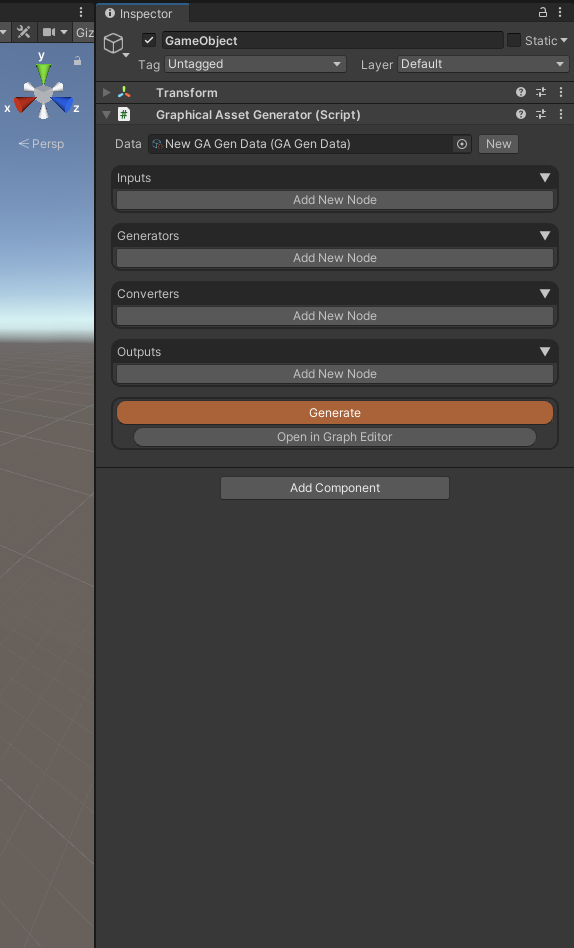}
  \\ Creating a new graphical asset generator 
\endminipage\hfill
\minipage{0.195\textwidth}
\centering
  \includegraphics[width=\linewidth]{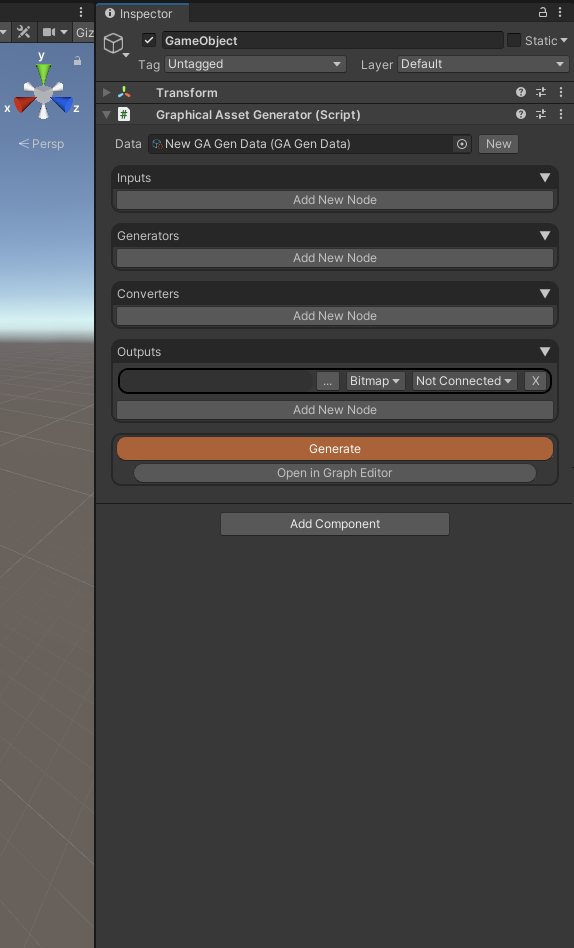}
    \\ Defining target asset type and format 
    \vspace{.4cm}
\endminipage\hfill
\minipage{0.195\textwidth}
\centering
  \includegraphics[width=\linewidth]{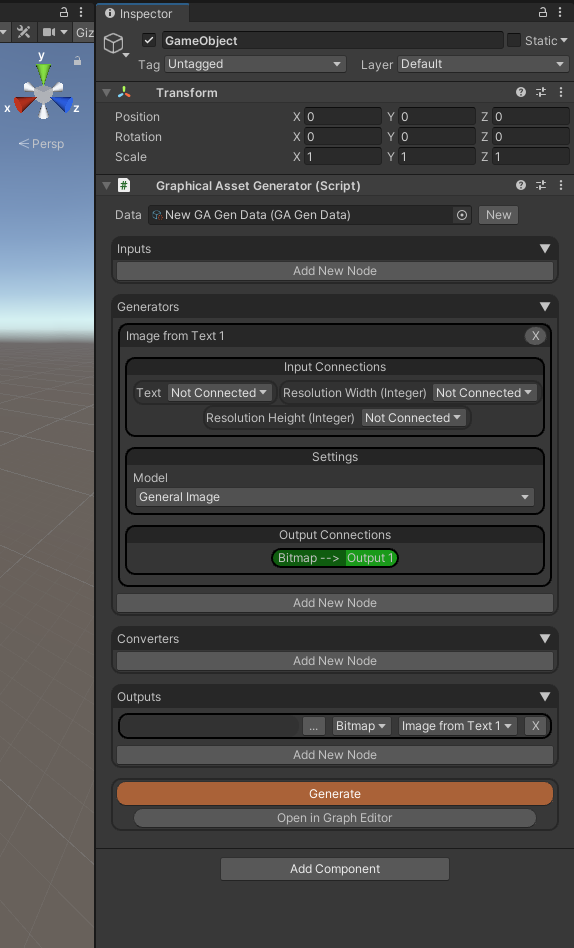}
  \\ Assigning the technique 
  \vspace{.4cm}
\endminipage\hfill
\minipage{0.195\textwidth}
\centering
  \includegraphics[width=\linewidth]{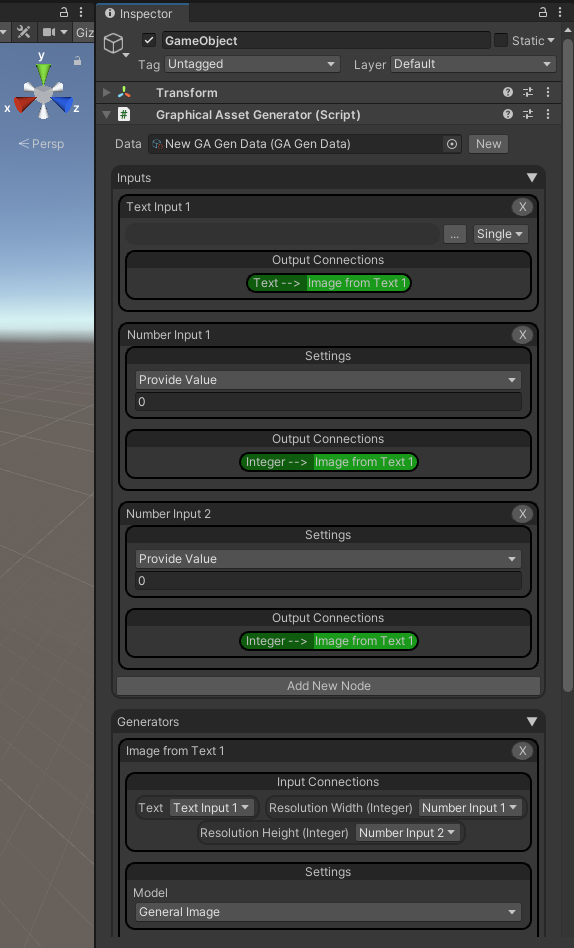}
  \\ Assigning input sources 
  \vspace{.4cm}
\endminipage\hfill
\minipage{0.195\textwidth}
\centering
  \includegraphics[width=\linewidth]{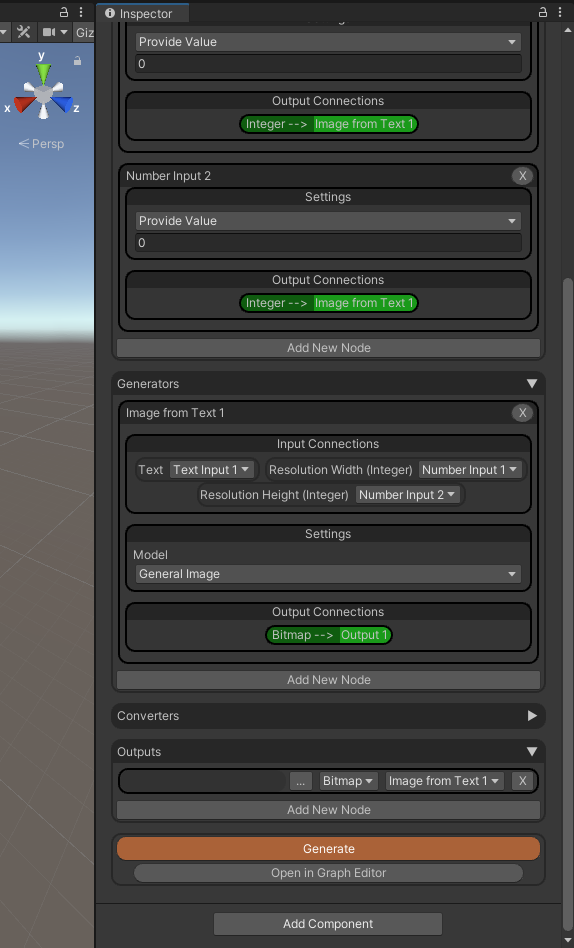}
  \\ Assigning output sources 
  \vspace{.4cm}
\endminipage
\vspace{.25cm}
\minipage{\textwidth}
    \minipage{\textwidth}
    \centering
    \textit{Integrated in an editor inspector view interface (M\textsubscript{3})}
    \endminipage
\endminipage
\end{figure*}

The three mock-up interfaces represent \textit{stand-alone wizard}, \textit{editor window integrated} and \textit{editor inspector integrated} implementations of this system. Figure \ref{fig:mockupSteps} shows an example usage process implemented in each of the three mock-ups. This process involves the configuration of a generator that takes text as input, and outputs bitmap images.

\subsection{\textit{M\textsubscript{1}: Stand-alone wizard}}
The first mock-up (\textit{M\textsubscript{1}}) represents a stand-alone wizard application in which the user answers a series of questions that determine the configuration of the generator. This mock-up was created using the Electron framework which packages designs built for web into an executable form. In addition, this mock-up was hosted as a web-page for users that did not have system permissions to run the application version. 

A software wizard is an application designed to guide a user through a series of steps \citep{nngroup_wizards}. These steps can be dependent on choices made in previous steps, thus allowing for branching behaviour. This can be used to match the logical process for GaGeTx, with information from each procedure feeding into later procedures.

The goal of this UI was to directly present the decision process proposed within the GaGeTx framework. The UI presents relevant multiple-choice questions that drill-down on a configuration based on the user's needs. The options are dynamic, such that certain answers limit the choices for certain proceeding questions. The result is a save-able generator configuration that the user can edit and return to. A breadcrumb navigation system allows the user to return to earlier choices and make changes as needed.

As shown in figure \ref{fig:mockupSteps}, this mock-up introduces the process, asks the user what type of asset they need to generate, the format, and then the input type. The choice of input then determines the possible techniques, in this example, there is only one choice so the user does not receive a selection. This is because there is only one technique for image generation via text. The user then arrives at a screen in which they choose input and output data paths, set the number of assets to generate, and begin the generation process.

\subsection{\textit{M\textsubscript{2}: Integrated in an editor window}}
Mock-up \textit{M\textsubscript{2}} represents an interface that would be integrated within existing game development software, in this case, the Unity engine. This uses Unity's experimental graph view API to render a node-based interface for designing generator configurations. This is incorporated as a separate sub-window of the Unity editor application.

As with the other mock-ups, this mock-up aims to present the GaGeTx framework decision process. For \textit{M\textsubscript{2}} the order in which decisions can be taken is up to the user. The UI provides four categories of node: inputs, generators, convertors, and outputs. The user can chain nodes from these categories to produce their desired configuration. Each node contains its own parameters that the user may adjust. These configurations can be saved as asset files within the Unity project. This UI also includes two modes: \textit{Generate}, and \textit{Train}. The set of nodes within the graph view persist between both modes, but connections between them can vary between the two. In generate mode you define outputs under a single output node. You then chain together input, generator and convertor nodes with the goal of connecting the final result to an output. In train mode the flow between nodes is indicative of a training pipeline. Alternative input paths can be provided as training data, labeling nodes can be inserted for labeling input data and output paths are disabled. The goal is to present this as a tool in which you design a pipeline, then fine-tune its inner workings. As this UI is an editor window, it is suited to \textit{offline} content generation as it does not directly connect with the active game scene. Instead, it directs outputs to user chosen folder paths from which the user can browse, refine or export results.

\textit{M\textsubscript{2}} is shown in figure \ref{fig:mockupSteps}. Here, the user creates a new bitmap path in the output node, creates an "image from text" node, and then creates an input path node while defining image size parameters.

\subsection{\textit{M\textsubscript{3}: Integrated in an editor inspector view}}
Mock-up \textit{M\textsubscript{3}}, also built in Unity, presents the same node based system as \textit{M\textsubscript{2}}, but is instead incorporated within the editor inspector window. This means that the interface is attached to an entity within the game scene, that the user must have selected in order to edit. This has the same functionality as \textit{M\textsubscript{2}}, however it is presented in a vertical, linear format.

The same categories available in \textit{M\textsubscript{2}} are shown as collapsible sections in \textit{M\textsubscript{3}}. Nodes are added under these categories using the respective "Add New Node" buttons which produce a relative list of nodes to choose from. To connect these nodes the user must select connections at the top of each node UI. An indicator at the bottom of each node will turn green if the node has been connected to, and which node this is. In the Unity Editor, inspector UI applies to a specific entity within a game scene. In a full implementation, this would give this version of the tool the ability to instantly load generated content within the game scene at a position of the user's choosing, facilitating rapid prototyping, variation, or runtime generation.

In figure \ref{fig:mockupSteps}, \textit{M\textsubscript{3}} presents a vertical list of categories which the user can add to. The user adds an output node with the format of "bitmap", adds an "image from text" generator node, and creates the necessary input nodes. These nodes are then bound to each other using drop-downs in the UI.

\section{Experiment Results}\label{sec:results}
All quantitative data from the questionnaires was analysed univariately, extracting average scores from Likert scale and single-choice questions, and frequency distributions for multiple choice questions. For this, one-sample Kolmogorov-Smirnov (K-S) \citep{KSSubcommand} tests were used to determine whether results were normally distributed. Wilcoxon signed-rank tests \citep{wilcoxon1992individual}, and one-way-ANOVA \citep{ross2017one} were conducted for each question, comparing the results from the three questionnaires. These results answer hypotheses H1 to H7, with respect to variance between mock-ups.
RH1 to RH3 are assessed via linear regression and independent samples T-Tests \citep{Independent-samples_T-test_2021},  observing the relationship between demographic and preference data, and an inductive thematic analysis was conducted on the notes from the semi-structured interview, answering RQ8 and RQ9.
During the data collection period, a total of 16 participants tested the mock-ups and completed the questionnaire, 4 of which also completed informal follow-up interviews. The initial mock-up testing phase took participants an average of 63 minutes to complete. 

K-S tests were performed for each question to determine if the distributions are normal. We find RQ7 (the SUS results) and RQ2 frequency of options picked (Table \ref{tab:ks}), to have normal distributions with statistical certainty (P $ > 0.05$). All other questions meet the null hypothesis (P $< 0.05$).

\begin{table}[]\caption{One-sample Kolmogorov-Smirnov test results for RQ1 - RQ7, presenting the mean, standard deviation (SD), and significance scores.}\label{tab:ks}
\centering
\begin{tabular}{lllll}
\multicolumn{2}{l}{Research   Question}                    & Mean  & SD      & Sig.\\ \toprule
\multicolumn{2}{l}{RQ1: (0-4)}                             & 2.72  & .669    & .001                     \\\midrule
\multirow{8}{*}{RQ2:} & RQ2.1   (0-1)                      & .67   & .485    & .000 \\
                      & RQ2.2   (0-1)                      & .72   & .461    & .000 \\
                      & RQ2.3   (0-1)                      & .61   & .502    & .000 \\
                      & RQ2.4   (0-1)                      & .17   & .383    & .000 \\
                      & RQ2.5   (0-1)                      & .06   & .236    & .000 \\
                      & RQ2.6   (0-1)                      & .17   & .383    & .000 \\
                      & RQ2.7   (0-1)                      & .39   & .502    & .000 \\
                      & \shortstack[l]{RQ2 \\ \small{Frequency   of options picked.}} & 2.78  & 1.629   & .194                     \\ \midrule
\multicolumn{2}{l}{RQ3: (0-1)}                             & .17   & .383    & .000                     \\ \midrule
\multicolumn{2}{l}{RQ4: (0-6)}                             & 1.67  & .686    & .001                     \\\midrule
\multicolumn{2}{l}{RQ5: (0-1)}                             & .78   & .428    & .000                     \\\midrule
\multicolumn{2}{l}{RQ6: (0-4)}                             & 3.50  & .985    & .004                     \\\midrule
\multicolumn{2}{l}{RQ7:   (0-100)}                         & 78.75 & 13.4287 & .131                     \\\bottomrule
\end{tabular}
\end{table}

\subsection{Descriptive statistics}
Overall, responses to RQ1 (usefulness) were moderately positive ($\mu = 2.72, \sigma = .669$). Of the options in RQ2, the most popular choices ($\mu > .5$) were RQ2.1 (generating inspiration), RQ2.2 (exploring ideas) and RQ2.3 (creating placeholder assets). The number of options picked by participants was between 2 and 3 ($\mu = 2.78, \sigma = 1.629$). For RQ3, participants show a clear preference for a high volume of lower quality assets over a lower volume of high quality assets ($\mu = .17, \sigma = .383$). RQ4 responses show an acceptable generation time for a single asset to be between 1 and 10 minutes ($\mu = 1.67, \sigma = .686$). For comparison, existing tools such as Didimo \citep{didimo} can take 5-10 minutes to produce a result, so this may be an expectation based on what is already available. RQ5 responses point toward preference for a tool integrated within a game-engine or editor ($\mu = .78, \sigma = .428$). For RQ6, results show that participants considered the ability to configure or modify the tools important ($\mu = 3.50, \sigma = .985$). The overall usability (RQ7) across all mock-ups was considered good ($\mu = 78.75, \sigma = 13.4287$).

\subsection{Tool Preference}
In order to determine the statistical significance in the difference in preference between the three mock-ups, ANOVA and Wilcoxon tests were conducted for the research questions RQ1 to RQ7. As seen in Table \ref{tab:oneWayAnova}, none of the questions presented statistically significant difference in preference between the mock-ups when tested using one-way ANOVA. This is confirmed in Table \ref{tab:wilcoxonSignedRanks}, where the Wilcoxon tests similarly report no statistically significant difference. Therefore, there is no significant difference in preference between the three mock-ups.

Beyond this, there is a slight leaning in the selection of RQ2.7 (user made content) for M2 and M3 over M1 (Z $ = -1.414$). RQ2.6 (assets from scratch) is chosen more for M1 than M3 (Z $ = -1.414$), and slightly more than M2 (Z $ = -1.000$). RQ2.4 (variations of complete assets) is chosen more for M2 than M3 (Z $ = 1.1414$) and slightly more than M1 (Z $ = -1.000$). Overall these ANOVA and Wilcoxon Test results suggest a slight separation between stand-alone (M1) and integrated (M2 and M3) mockups.

Further ANOVA and Wilcoxon tests were conducted between the integrated and stand-alone solutions for RQ1 (usefulness) and RQ5 (preferred as stand-alone or integrated), shown in table \ref{tab:anovaWilcoxonSvsI}. These also present non significant differences, though we note a slightly higher mean RQ1 score for integrated mockups, and a leaning towards preference for integrated solutions overall (RQ5 $\mu= .78$).

\begin{table*}[] \caption{One-way ANOVA significance scores for RQ1 to RQ7, alongside means and standard deviations per mock-up.}\label{tab:oneWayAnova}
\centering
\begin{tabular}{llll|ll|ll|lll}
\multicolumn{2}{l}{\multirow{2}{*}{Research   Question}} & \multicolumn{2}{c}{M1} & \multicolumn{2}{c}{M2} & \multicolumn{2}{c}{M3} & \multicolumn{3}{c}{Overall} \\
\multicolumn{2}{l}{}                                     & Mean      & SD         & Mean      & SD         & Mean      & SD         &     Mean                       &               SD  &   Sig. \\ \toprule
\multicolumn{2}{l}{RQ1: (0-4)}                           & 2.50      & .548       & 2.83      & .753       & 2.83      & .753       & 2.72                       & .669                     & .637                  \\ \midrule
\multirow{7}{*}{RQ2:}           & RQ2.1   (0-1)          & .50       & .548       & .83       & .408       & .67       & .516       & .67                        & .485                     & .521                  \\
                                & RQ2.2   (0-1)          & .67       & .516       & .83       & .408       & .67       & .516       & .72                        & .461                     & .791                  \\
                                & RQ2.3   (0-1)          & .50       & .548       & .67       & .516       & .67       & .516       & .61                        & .502                     & .821                  \\
                                & RQ2.4   (0-1)          & .17       & .408       & .33       & .516       & .00       & .000       & .17                        & .385                     & .342                  \\
                                & RQ2.5   (0-1)          & .00       & .000       & .17       & .408       & .00       & .000       & .06                        & .236                     & .391                  \\
                                & RQ2.6   (0-1)          & .33       & .516       & .17       & .408       & .00       & .000       & .17                        & .383                     & .342                  \\
                                & RQ2.7   (0-1)          & .17       & .408       & .50       & .548       & .50       & .548       & .39                        & .502                     & .439                  \\\midrule
\multicolumn{2}{l}{RQ3: (0-1)}                           & .17       & .408       & .17       & .408       & .17       & .408       & .17                        & .383                     & 1.000                 \\\midrule
\multicolumn{2}{l}{RQ4: (0-6)}                           & 1.67      & .816       & 1.67      & .816       & 1.67      & .516       & 1.67                       & .686                     & 1.000                 \\\midrule
\multicolumn{2}{l}{RQ5: (0-1)}                           & .67       & .516       & .83       & .408       & .83       & .408       & .78                        & .428                     & .761                  \\\midrule
\multicolumn{2}{l}{RQ6: (0-4)}                           & 3.50      & 1.378      & 3.50      & .548       & 3.50      & 1.049      & 3.50                       & .985                     & 1.000                 \\\midrule
\multicolumn{2}{l}{RQ7: (0-100)}                         & 80.00     & 12.649     & 78.75     & 10.694     & 77.50     & 18.303     & 78.75                      & 13.429                   & .995                  \\\bottomrule
\end{tabular}
\end{table*}

\begin{table*}[]\caption{Z-values and significance (P-values) for Wilcoxon Signed Ranks Test between each mock-up, for each question.}\label{tab:wilcoxonSignedRanks}
\centering
\begin{tabular}{llll|ll|ll}
\multicolumn{2}{c}{\multirow{2}{*}{}} & \multicolumn{2}{c}{M2 -\textgreater   M1} & \multicolumn{2}{c}{M3 -\textgreater   M1} & \multicolumn{2}{c}{M3 -\textgreater   M2} \\
\multicolumn{2}{l}{}& Z& Sig.      & Z           & Sig.      & Z           & Sig.     \\ \toprule
\multicolumn{2}{l}{RQ1}                                  & -1.414      & .157                        & -1.414      & .157                        & .000        & 1.00\\ \midrule
\multirow{8}{*}{RQ2}        & RQ2.1                      & -1.000      & .317                        & .000        & 1.00                       & -1.00& .317                        \\
                            & RQ2.2                      & -1.000      & .317                        & .000        & 1.00                       & -1.000      & .317                        \\
                            & RQ2.3                      & -1.000      & .317                        & .000        & -1.00                      & .317        & 1.00                       \\
                            & RQ2.4                      & -1.000      & .317                        & -1.00      & .317                        & 1.1414      & .157                        \\
                            & RQ2.5                      & -1.000      & .317                        & .000        & 1.00                       & -1.000      & .317                        \\
                            & RQ2.6                      & -1.000      & .317                        & -1.414      & .157                        & -1.00      & .317                        \\
                            & RQ2.7                      & -1.414      & .157                        & -1.414      & .157                        & .000        & 1.00                       \\
                            & Total        & -1.342      & .180                        & -1.00      & .317                        & -.816       & .414                        \\\midrule
\multicolumn{2}{l}{RQ3}                                  & .000        & 1.000                       & .000        & 1.00& .000        & 1.00\\ \midrule
\multicolumn{2}{l}{RQ4}                                  & .000        & 1.000                       & .000        & 1.00& .000        & 1.00\\ \midrule
\multicolumn{2}{l}{RQ5}                                  & -1.000      & .317                        & -1.000      & .317                        & .000        & 1.00\\\midrule
\multicolumn{2}{l}{RQ6}                                  & .000        & 1.000                       & .000        & 1.00& .000        & 1.00\\\midrule
\multicolumn{2}{l}{RQ7}                                  & -.406       & .684                        & -.184       & .854                        & .713        & .713   \\     \bottomrule               
\end{tabular}
\end{table*}

\begin{table*}[] \caption{One-way ANOVA and Wilcoxon Signed Ranks tests comparing stand-alone mock-up results (M1) to integrated mock-up results (M2 \& M3), for relevant research questions.}\label{tab:anovaWilcoxonSvsI}
\centering
\begin{tabular}{lll|ll|lll|ll}
\multirow{3}{*}{Research   Question} & \multicolumn{7}{c}{One-way   ANOVA}                                                                                                     & \multicolumn{2}{c}{Wilcoxon   Signed Ranks}     \\
                                     & \multicolumn{2}{c}{M1} & \multicolumn{2}{c}{M2 \&   M3} & \multicolumn{3}{c}{Overall}  & \multicolumn{2}{c}{M2 \&   M3 -\textgreater M1} \\
                                     & Mean  & SD  & Mean & SD & Mean & SD & Sig. & Z   & Sig. \\ \hline
RQ1                                  & 2.50       & .548      & 2.83           & .718          & 2.72                       & .669                     & .334                  & -1.414         & .157                           \\
RQ5                                  & .67        & .516      & .83            & .389          & .78                        & .428                     & .453                  & -1.000         & .317                          
\end{tabular}
\end{table*}

\subsection{Demographic impact}
13 of the 16 participants were male, and 13 participants were under the age of 25. Additionally, 10 participants had under 1 year of experience in professional game development, while 3 had between 1 and 4 years, and 3 had between 5 and 9 years. 12 participants typically work solo, and 4 participants typically work in a team size of 11-25. The most common typical roles of participants were artist and designer (50\%), followed by technical artist and programmer (31.25\%).
To examine impact, multiple regression with a confidence interval of 95.0\% was conducted for each single choice scale question (RQ1, RQ4, RQ6 and RQ7), against participant demographic variables. Table \ref{tab:lrResultsSingle} presents the results of the multiple regression for single-choice scale questions, showing their correlation with the four single-choice demographic questions. As RQ3 and RQ5 are binary questions, Independent-Samples T-tests were instead conducted, comparing demographic answers between two groups, where the the groups are defined by the binary answer to the option (chosen or not chosen). This approach was also taken with RQ2, where multiple choice answers were binary, due to being tick-boxes. These results are presented in table \ref{tab:ttestResultBinary}.

RQ2.5 "Creating assets from existing (complete) designs" (shown in grey table \ref{tab:ttestResultBinary}), was only selected once out of all samples. As a result there was not enough variance to report a T-test result. The RQ3 gender T-test could not be completed due to standard deviations of both groups being zero, this is also shown in grey.

The ANOVA on the linear regressions of RQ1 (usefulness), RQ6 (importance of configurability) and RQ7 (SUS results) in table \ref{tab:lrResultsSingle}, are statistically significant (P $< 0.05$). Of these, RQ7 has a strong correlation with the four tested demographic questions ($Adjusted R Square = .671$), while RQ1 ($Adjusted R Square = .305$) and RQ6 ($Adjusted R Square = .428$) show low correlation. For RQ7 age, years of professional experience and size of team have high negative correlations. That is, the older and more experienced participants, and those that work in larger teams, gave lower SUS ratings.
A moderate negative correlation for age, years of experience and size of team is also observed in results for RQ6 and RQ1. Additionally there is an insignificant and low degree of correlation between gender and results of RQ6 and RQ1.

T-test results in table \ref{tab:ttestResultBinary} show similar levels of impact for the four demographic categories. Gender's impact is insignificant in all cases, while years of professional experience and size of team are significant for RQ2.1, RQ2.2 and RQ3. Positive T values for RQ2.1 and RQ2.2 show that those with more professional experience and larger team sizes choose these options (generating inspiration, and exploring ideas) more frequently. More years of professional experience ($P = .014, T = -2.806$) and a larger size of team ($P = .000, T = -7.483$) both point toward the choice of volume over quality, though a near unanimous selection of volume over quality was made by participants.

While years of professional experience and age are intrinsically linked, we find that they do not have equivalent impact in most cases. This suggests that the two are decoupled. Age does not have a significant impact in any of the T-test results, while years of professional experience does. The level of experience and size of team both correlate with lower ratings (RQ1/RQ7) and interest in using the tools for generating inspiration and exploring ideas (RQ2.1/RQ2.2). This is corroborated by the near unanimous selection of volume over quality. We can conclude that experienced users prefer graphical asset generation in early ideation/inspiration stages of creation. This is also reflected in the choices overall for RQ2.1 ($\mu = .67$), and RQ2.2 ($\mu = .72$).

\begin{table*}[] \caption{Linear regression results for RQ1, RQ4, RQ6 and RQ7 (all single-choice questions scale), reporting Pearson correlation \citep{pearsoncommand}, R Square \citep{rsquare}, adjusted R Square and ANOVA p-value for each single-choice demographic question. (*) \textit{Experience} stands for \textit{Years of professional experience}}\label{tab:lrResultsSingle}
\centering
\begin{tabular}{llll|ll|l}
\multicolumn{2}{l}{\multirow{2}{*}{Research   Question}}  & \multicolumn{2}{c}{Correlation} &  \multicolumn{2}{c}{Model Summary} & ANOVA                  \\
\multicolumn{2}{l}{}                                      & Correlation    & Sig.   & R Square               & Adjusted R Square      & Sig.                   \\\hline
\multirow{4}{*}{RQ1} & Age                                & -.497                  & .018   & \multirow{4}{*}{.387}  & \multirow{4}{*}{.305}  & \multirow{4}{*}{.026}  \\
                     & Gender                             & -.267                  & .142   &                        &                        &                        \\
                     &Experience*                         & -.615                  & .003   &                        &                        &                        \\
                     & Size of team                       & -.604                  & .004   &                        &                        &                        \\ \midrule

\multirow{4}{*}{RQ4} & Age                                & -.447                  & .031   & \multirow{4}{*}{.302}  & \multirow{4}{*}{.209}  & \multirow{4}{*}{.067}  \\
                     & Gender                             & -.224                  & .186   &                        &                        &                        \\
                     & Experience* & -.546                  & .010   &                        &                        &                        \\
                     & Size   of team                     & -.530                  & .012   &                        &                        &                        \\ \midrule

\multirow{4}{*}{RQ6} & Age                                & -.701                  & .001   & \multirow{4}{*}{.495}  & \multirow{4}{*}{.428}  & \multirow{4}{*}{.006}  \\
                     & Gender                             & .078                   & .379   &                        &                        &                        \\
                     & Experience* & -.646                  & .002   &                        &                        &                        \\
                     & Size   of team                     & -.492                  & .019   &                        &                        &                        \\ \midrule
\multirow{4}{*}{RQ7} & Age                                & -.614                  & .003   & \multirow{4}{*}{.710}  & \multirow{4}{*}{.671}  & \multirow{4}{*}{.000}  \\
                     & Gender                             & -.443                  & .033   &                        &                        &                        \\
                     & Experience* & -.815                  & .000   &                        &                        &                        \\
                     & Size   of team                     & -.835                  & .000   &                        &                        &                       \\\bottomrule
\end{tabular}
\end{table*}

\setlength{\tabcolsep}{4pt}
\begin{table*}[ht] \caption{Independent-Samples T Test results for RQ2, RQ3 and RQ5 (all binary-choice questions), reporting t-statistic (t) and p-value (sig.). (*) \textit{Experience} stands for \textit{Years of professional experience}}\label{tab:ttestResultBinary}
\centering
\begin{tabular}{c|cccc|cccc|cccc|cccc|c|}
\toprule
\multirow{2}{*}{\shortstack{Research \\Question}} & \multicolumn{4}{c|}{RQ2.1} & \multicolumn{4}{c|}{RQ2.2} & \multicolumn{4}{c|}{RQ2.3} & \multicolumn{4}{c|}{RQ2.4} & \multicolumn{1}{c|}{RQ2.5}\\
&\rotatebox[origin=l]{90}{Age}&\rotatebox[origin=l]{90}{Gender}&\rotatebox[origin=l]{90}{Experience*}&\rotatebox[origin=l]{90}{Size of team}&\rotatebox[origin=l]{90}{Age}&\rotatebox[origin=l]{90}{Gender}&\rotatebox[origin=l]{90}{Experience*}&\rotatebox[origin=l]{90}{Size of team}&\rotatebox[origin=l]{90}{Age}&\rotatebox[origin=l]{90}{Gender}&\rotatebox[origin=l]{90}{Experience*}&\rotatebox[origin=l]{90}{Size of team}&\rotatebox[origin=l]{90}{Age}&\rotatebox[origin=l]{90}{Gender}&\rotatebox[origin=l]{90}{Experience*}&\rotatebox[origin=l]{90}{Size of team}&\rotatebox[origin=l]{90}{\cellcolor{gray!25}{}}\\
T&2.24&1.10&3.64&4.54&2.45&1.26&5.93&7.30&-1.94&.95&-1.05&-.32&1.87&-.82&.39&.00&\cellcolor{gray!25}{}\\
Sig.& .08& .31& .01& .00& .07& .27& .00& .00& .08& .37& .31& .75& .08& .43& .70& 1.00&\cellcolor{gray!25}{}\\
\toprule
\multirow{2}{*}{\shortstack{Research \\Question}}& \multicolumn{4}{c|}{RQ2.6} & \multicolumn{4}{c|}{RQ2.7} & \multicolumn{4}{c|}{RQ3} & \multicolumn{4}{c|}{RQ5}\\
&\rotatebox[origin=l]{90}{Age}&\rotatebox[origin=l]{90}{Gender}&\rotatebox[origin=l]{90}{Experience*}&\rotatebox[origin=l]{90}{Size of team}&\rotatebox[origin=l]{90}{Age}&\rotatebox[origin=l]{90}{Gender}&\rotatebox[origin=l]{90}{Experience*}&\rotatebox[origin=l]{90}{Size of team}&\rotatebox[origin=l]{90}{Age}&\rotatebox[origin=l]{90}{Gender}&\rotatebox[origin=l]{90}{Experience*}&\rotatebox[origin=l]{90}{Size of team}&\rotatebox[origin=l]{90}{Age}&\rotatebox[origin=l]{90}{Gender}&\rotatebox[origin=l]{90}{Experience*}&\rotatebox[origin=l]{90}{Size of team}\\
T&1.87&-1.77&-.39&-1.33&1.94&-.95&1.05&.32&1.87&\cellcolor{gray!25}{}&-2.81&-7.48&1.44&-1.88&1.07&.77\\
Sig.& .08 & .21& .70& .20& .08& .37& .31& .75& .08&\cellcolor{gray!25}{}& .01& .00& .24& .08& .35& .45\\
\end{tabular}
\end{table*}

\subsection{Interview Results} 
During the interviews, the participants generally confirmed the results of the questionnaires. There is no clear preference between the three mock-ups but they emphasised compatibility with their existing tools. One participant mentioned 
that each mock-up was fine, as long as they didn't need a converter or third tool to get the assets into their game. Most participants confirmed that they would prefer to use the tool for ideation and therefore prefer speed and volume over quality. 

Overall, the participants were very supportive and excited about content generation tools and use of AI in their pipelines. The main concern they had was regarding the ease of use and prior knowledge of AI or PCG needed to be able to use the tool effectively. for example a participant pointed that whilst they would not necessarily expect tools such as DALL-E and Mid-Journey to be integral part of game design and development but did like the idea of a middle ground so that they can have control but also use the tool relatively easily. There was not a clear preference towards the type of assets they would prefer to generate with such a tool. Most participants mentioned a wide variety of asset types, based on their current project, but also highlighted that any one type would also be useful. Furthermore, with regard to the format used to store these assets, participants expressed that as long as the assets were in a standard, readable format for their game engine of choice, there were no concerns.

With regard to usability, one participant noted that they found it frustrating that they could not undo and redo their changes within the Unity integrated tools. They stated that this was a feature that they expected, considering that the tool appeared to be part of the engine interface that they were familiar with.

\subsection{Findings}
For H1, given answers to RQ5, we find that users prefer an integrated solution over a stand-alone implementation, with a mean value of .78 ($\sigma$ = .428), where 1 represents an ‘integrated solution’. This is confirmed by RQ1, in which the integrated solutions scored higher. With regards to H2, results of RQ1 and RQ6.1 suggest no significant preference between window and inspector integrated interfaces.
H4 is confirmed with a preference for volume over quality across the board, with a mean value of .17 ($\sigma$ = .383), where 0 represents ‘volume over quality’. Regarding H5 (RQ4), a mean selection of 1.67 ($\sigma$ = .686) shows a preference for a maximum generation time between 1 and 10 minutes for a single asset, rejecting this hypothesis. For H6, a mean score of 3.5 ($\sigma$ = .985) on a 5-point Likert scale, RQ6, suggests a moderate preference for the ability to configure or modify the tool. Answers to RQ7 present an overall mean SUS  score of 78.75 ($\sigma$ = 13.429), therefore confirming H7. The individual mean SUS scores for M1, M2 and M3 were 80.00 ($\sigma$ = 12.649), 78.75 ($\sigma$ = 10.694), and 77.50 ($\sigma$ = 18.303) respectively. Following the adjective rating system provided by Bangor et al. \citep{bangor2009determining}, the mean overall score can be classed as "Good" (78.75). The next sections will discuss these findings and propose a set of heuristics, as well as suggestions for future research.

\begin{figure*}[ht] 
\centering
\includegraphics[width=1\linewidth]{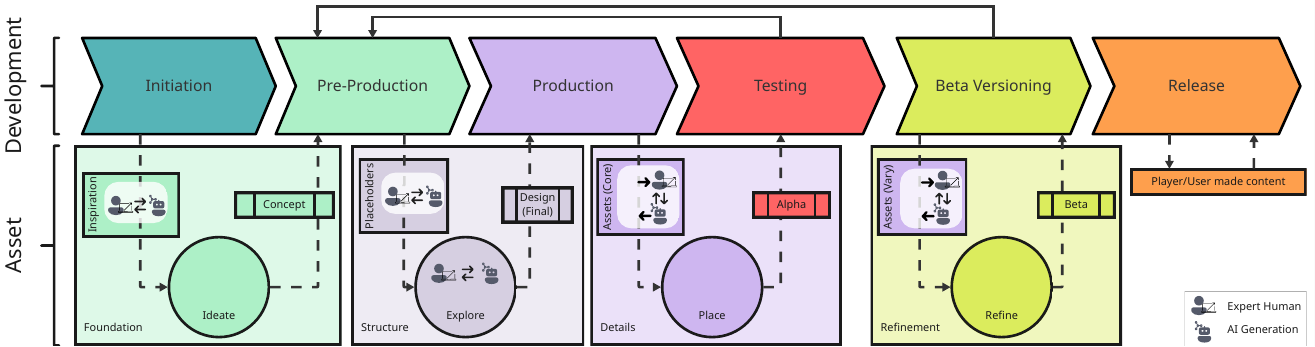}
\caption{The game design and development pipeline based on the GDLC of Ramadan and Widyani \citep{Ramadan2013} and expanded to include graphical asset findings.}\label{fig:devpipeline}
\end{figure*}

\section{Discussion and Comparison of Results}\label{sec:comparison}
This study has found that there is a preference for all UIs to be integrated into game-engines/editors, and for the system to generate larger amounts of lower quality assets as opposed to smaller amounts of higher quality assets. This is also corroborated by a preference for using such a system in early stages of development to generate inspiration, explore ideas and create placeholders. The ability to configure or modify such a tool is also considered important. The maximum acceptable time scale for generating a single asset is between 1 and 10 minutes. This is not surprising, as existing tools such as Didimo \citep{didimo} can take 5-10 minutes to generate assets, especially at higher qualities. The other findings suggest that a much faster generation speed would be ideal, in order to provide larger volumes of assets for inspiration and ideation. There is however no statistically significant difference in preference between a stand-alone multi-choice based UI, integrated graph-view UI and integrated inspector UI for generating graphical assets.

\begin{figure}[ht]
    \centering
    \includegraphics[width=.5\linewidth]{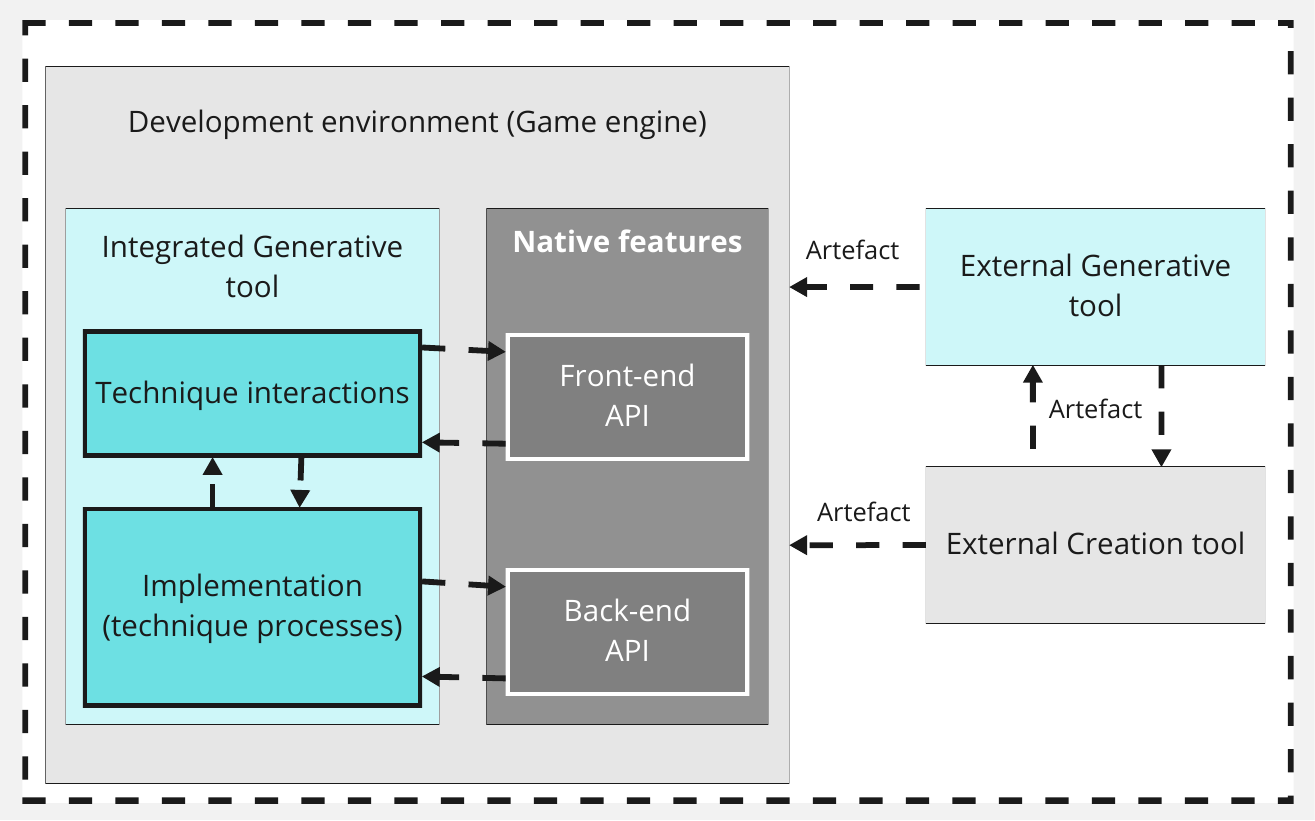}
    \caption{Graphical asset creation within the game production framework.}
    \label{fig:gametool-schematic}
\end{figure}

Users were interested in using the tool for inspiration and prototyping, and favoured integration, and thus compatibility with their chosen game engine. Additionally, when observing the characteristics of the three preferred usages according to the characteristics shown in figure \ref{fig:pipelinecat}, in general, users prefer generators for work in an \textit{offline} context, with an expectation for lower quality and to augment, rather than replace an existing process in their pipeline.

Given the findings regarding pipeline preferences, we refine the graphical asset GDLC to include the \textit{offline} generative tool applications, shown in figure \ref{fig:devpipeline}. Here, the pipeline applications \textit{creating assets from existing designs} and \textit{creating assets from scratch} both apply under the \textit{Assets (core)} step and have thus been combined. We also show the human-computer interaction for each of these steps with early pipeline tasks: inspiration, placeholders and explore, employing an iterative shared initiative approach, where both human and AI provide equal input; and late pipeline tasks: Assets (core) and Assets (vary), taking a human-driven approach, in which designs are already well defined and the goal is to implement them in a more controlled and selective manner. This addition formally situates the well-corroborated preference for applying generative tools in the early stages of the pipeline.

Our results align with and expand upon existing research findings regarding game design and development tools \citep{walton2021evaluating, kasurinen2013game, Ling2024sketchar, colado2023using, Togelius2011, lai2020towards}, addressing the gaps in previous research. Namely, previous work has suggested that designers and developers value variety in outputs \citep{walton2021evaluating}, which corroborates with the usage of generative methods as explorative tools \citep{Ling2024sketchar, colado2023using, kasurinen2013game}. Our results not only reflect this need and preference, but identify the exact use cases designers and developers target within the pipeline, providing further context to this requirement and allowing tools to cater to those specific needs. 

Although previous research demonstrates an interest in using these tools in pipelines \citep{Ling2024sketchar, colado2023using}, they do not clarify how and where in the pipeline GAG tools would be suitable. Our findings show that they are much more preferred in the early stages. Furthermore, we expand on research that suggests designers and developers prefer well integrated configurable tools \citep{kasurinen2013game} through direct quantitative results and interview feedback emphasising compatibility with their tools. Our results show that the style of interface only matters as far as it suits the interactions needed for the technique in use. To integrate these interactions into existing software the interface must be achieved through APIs, which also inform the design language of the tool. 

Previous works focus on high-level tool implications \citep{Togelius2011, walton2021evaluating}, but do not consider the practical aspects that can make or break the applicability of a tool. On the contrary, Lai et al. \citep{lai2020towards} identify a need to respect existing work processes. We elaborate on this by pinpointing the use of standardised data formats as an important requirement for generative tools in practice, as well as the need for design features that align with the host environment and thus meet user expectations. 


While existing work acknowledges that different balances of user-machine initiative can be useful in different situations \citep{Liapis2016, lai2020towards}, research has not yet ascertained how this balance maps onto the stages of a design and development pipeline. Through our findings we identify the appropriate flow of initiative at different stages of a graphical asset pipeline as presented in figure \ref{fig:devpipeline}. At earlier stages of the pipeline less user control is required, thus the tool can have a larger share of the initiative. Once development shifts into a production phase, the design choices have already been made, and thus the user takes a more prescriptive role over the tool. This helps to inform the design and integration of GAG tools, based on their intended stage in the pipeline.

The flow of graphical asset data in both stand-alone and integrated scenarios is presented in figure \ref{fig:gametool-schematic}. As shown, creative tools can either be integrated within the main development environment (game engine) or implemented externally as stand-alone programs. Users interact with these tools via user-interfaces, and in the case of integrated tools, the implementation itself can rely on the native features of the environment through application programming interfaces (API). Stand-alone external generative tools, as with other creation tools such as Photoshop \citep{adobephotoshop}, Blender \citep{BlenderFoundation2022} and  Visual Studio \citep{visualstudio} instead rely on outputting artefacts in standardised formats that most game engines support, such as FBX, OBJ, PNG and JPEG. Integrated tools may implement GAG technique \textit{interactions} and \textit{processes} using game engine UI APIs and native engine features via back end APIs respectively. 

\section{Heuristics for Generative Tools}\label{sec:heuristics}
\begin{figure}[ht] 
\centering
\includegraphics[width=.5\linewidth]{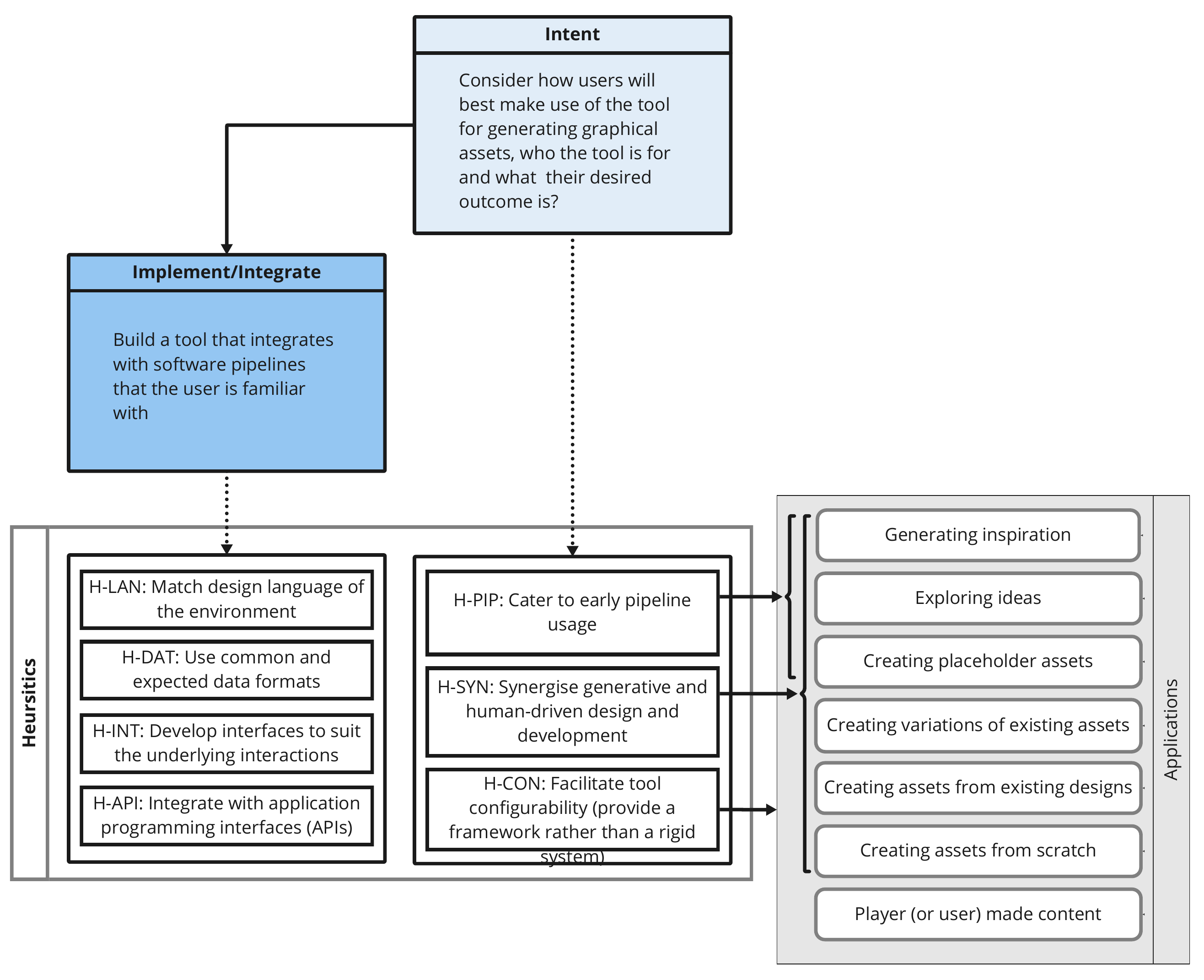}
\caption{Mapping the user considerations, intent and implementation/integration to the heuristics.}\label{fig:preference-outcomes}
\end{figure}

To improve the applicability of our findings to the design and integration of GAG tools, we have derived a set of actionable heuristics, shown in figure \ref{fig:preference-outcomes}, spanning both the user intent as well as the implementation and integration of GAGs in design/development pipelines. Table \ref{tab:heuristics} presents the mapping of these heuristics to the research questions.

\begin{table*}[ht]
\caption{Mapping of heuristics to research questions.}
\label{tab:heuristics}
\centering
\begin{tabular}{rll}
\multicolumn{2}{c}{Heuristics} & Research Question(s)\\ \toprule
H-CON & Facilitate tool configurability &RQ6\\\midrule
H-PIP & Cater to early pipeline usage & RQ2, RQ3, RQ4\\\midrule
H-SYN & Synergise generative and human-driven design and development &RQ2\\\midrule
H-LAN & Match the design language of the environment & RQ8\\\midrule
H-DAT & Use common and expected data formats & RQ8, RQ9\\\midrule
H-INT & Develop interfaces to suit the underlying interactions & RQ1, RQ7\\\midrule
H-API & Integrate with application programming interfaces (APIs) & RQ5, RQ8\\\bottomrule                
\end{tabular}
\end{table*}

The first consideration is user intent with regard to the usage application, what technique interaction and process is used and what the desired outcome is. When considering user intent, based on the findings, there are three heuristics:

\begin{itemize}
    \item \textit{H-PIP: Cater to early pipeline usage.}
    Users favour early pipeline generation for purposes of generating inspiration, exploring ideas, and creating placeholders, as presented in figure \ref{fig:preference-outcomes}. Furthermore, the system should aim for variety and volume of outputs rather than quality. This is because users prefer not to use the results in a finished product directly, but would rather use them as ideas to build on and refine.
    
    \item \textit{H-SYN: Synergise generative and human-driven design and development.}
    The preferred use cases are all supportive of a human-driven design and development process, and do not directly replace the final creation of assets. Rather, they give designers and developers ideas to expand on, break creative block or help to streamline more mundane or inconsequential tasks. This entails the majority of applications, excluding player made content as shown in figure \ref{fig:preference-outcomes}.

    \item \textit{H-CON: Facilitate tool configurability.}
    Game design and development is a creative endeavor and thus, new, unique forms of assets are part and parcel. A graphical asset generator that has a pre-defined style or rigid way of working becomes a tool with limited applicability. Configurability should be considered, regardless of the chosen application.
\end{itemize}

Once intent is established, and the GAG technique and approach is determined, the tool must be implemented and integrated within the game design and development pipeline. Here, there are four heuristics:

\begin{itemize}
    \item \textit{H-LAN: Match the design language of the environment.}
        The less the user has to learn on top of what they are accustomed to within their pipeline, the easier it is to incorporate the tool. If the interface of the tool provides functionality that other native features of the environment have, such as the ability to undo and redo changes, frustration or confusion can be avoided. This can be achieved through appropriate use of front-end UI APIs within the host software, as illustrated in figure \ref{fig:gametool-schematic}.
        
    \item \textit{H-DAT: Use common and expected data formats.} 
        To smoothly integrate the tool, whether it is integrated or stand-alone, the output artefacts must be in a format that the development environment can utilise. Typically these are ubiquitous data formats, such as OBJ or common proprietary formats that are supported due to popularity, such as Autodesk FBX. This is particularly important in the transfer of artefacts from and between external generative tools, creation tools and the users development environment of choice, as shown in figure \ref{fig:gametool-schematic}. 
        
    \item \textit{H-INT: Develop interfaces to suit the underlying interactions.} 
        Design an interface to best interact with the features of the tool, assuming that it provides good usability, there is no preference between step by step options, graph view windows and inspector editors.
        
    \item \textit{H-API: Integrate with application programming interfaces (APIs).} 
        The tool should be integrated as a part of an existing popular engine or editor, or at the very least should have full compatibility with said application. Technique interactions and processes must interact with each other while appropriately utilising the host game engine's front-end and back-end APIs, as shown in figure \ref{fig:gametool-schematic}.
\end{itemize}

\begin{figure*}[ht] 
\centering
\includegraphics[width=1\linewidth]{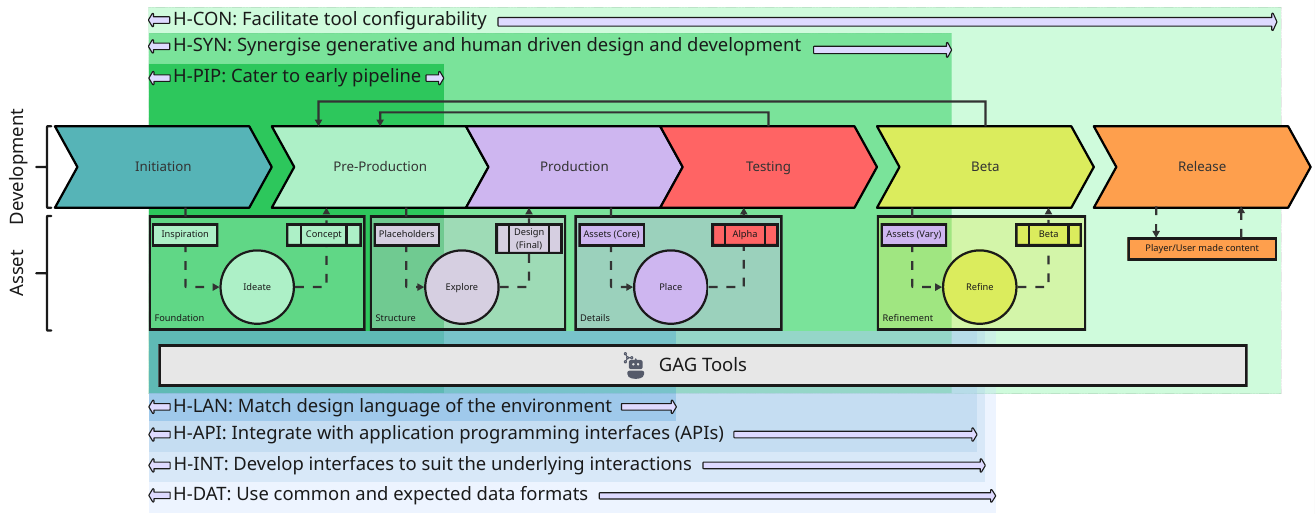}
\caption{Refined game asset design and development pipeline with heuristics resulting from research findings.}\label{fig:final-heuristic-pipeline}
\end{figure*}

\subsection{Theoretical and Practical Implications}

As GAG methods advance, their applications in games will continue to become more and more prevalent. Given our findings and resulting heuristics, the role of GAG tools in their present form within game pipelines is now clearer. With this, we adapt the existing GDLC framework \citep{Ramadan2013} to meet with the growing interest in GAG tools. We integrate these heuristics into the graphical asset GDLC, shown in figure \ref{fig:final-heuristic-pipeline}, illustrating their relevance at different stages of the pipeline.

As previously mentioned, whilst there is extensive work on the benefits and potentials of GAG tools in game design and development \citep{Li2021d, shen2020deepsketchhair, shen2021clipgen}, there are limited studies on their use and fit within a design and development pipeline \citep{Ling2024sketchar, colado2023using}. While the use of creative tools is well established \citep{kasurinen2013game, seidel2019designing}, the behaviour and preferences of game designers and developers with regard to GAG tools has not been explored. Hence, there are no existing heuristics or guidelines on how to effectively incorporate them into the design and development of a game.

This study has aimed to take the first steps in addressing this gap, and to contextualise the behaviour and preference of designers and developers within the larger scope of the development pipeline. The results demonstrate that designers and developers are clearly interested in incorporating graphical asset generation tools, and most are already using them in some capacity.  

There are a number of practical implications that arise from the study results. GAG tools should be developed for incorporation in the early stages of design and development, these tools should be integrated into existing design and development environment as much as possible, they should prioritise larger volumes of variations over quality, the users should be able to configure and have some control over the tools and the generated output artefacts, and finally the generated artefacts should comply with common data formats. These implications are derived from the behaviour and preferences of users as there is a strong preference towards using GAG tools in the early design stages for generating inspiration, exploring ideas and creating placeholders. They prefer to have a large volume of variations which they can later manipulate and improve to create their chosen assets in their intended quality. They demonstrate a strong interest in being able to configure GAG tools, compliance of outputs with the common data formats and control over the functionality and output formats of the generated artefacts.

Considering the specific practical implications discussed above, leads to a more high-level theoretical implication for the development of GAG tools. In order to efficiently take advantage of the opportunities provided by GAG tools, the practical needs and preferences of the game designers and developers are paramount. Whilst the current generic approach to AI driven content generation has provided powerful interesting tools, lack of empirical research in the behaviour and preferences of game designers and developers, has led to obstacles for incorporating them well into the design and development pipelines.

\section{Concluding Discussions}
This study has investigated the behaviour and preferences of game designers and developers towards the use of graphical asset generation tools, e.g. PCG, and other intelligent generative approaches. The study considers how these users interact with generative tools within the creation pipeline and takes into account the different contributing factors such as their experience, their role and the size of their respective teams. 
The results indicate that users prefer early design stage applications such as generating inspiration, exploring ideas and creating placeholders. As such, users prefer tools that can output larger volumes of artefacts at the cost of quality. No preference is shown for any one form of interface, instead, users express a need for smooth integration of GAG tools within their existing development environments. This includes complying with the design language of the host software (when integrated), presenting the necessary controls for the tool to function and outputting artefacts via common data formats. Users express a need for tool configurability, and find generation times of up to 10 minutes to be acceptable.
These findings have led to a set of heuristics for the use and integration of generative tools within game design and development pipelines, with considerations for the user's intended application.


\bibliographystyle{unsrtnat}

\end{document}